\documentclass[aps, prd, twocolumn, amsmath, floats,floatfix, superscriptaddress, nofootinbib,
showpacs]{revtex4}

\usepackage{amssymb}
\usepackage{amsmath}
\usepackage{verbatim}
\usepackage{mathrsfs}
\usepackage{amsfonts}
\usepackage{latexsym}
\usepackage{epsfig}
\usepackage{color}
\usepackage{graphicx,subfigure}
\usepackage{units}
\usepackage{envmath}
\usepackage{natbib}
\usepackage{ctable}
\usepackage{soul}

\begin{document}


\definecolor{orange}{rgb}{0.9,0.45,0}

\newcommand{\re}{\mbox{Re}}
\newcommand{\im}{\mbox{Im}}
\newcommand{\tf}[1]{\textcolor{red}{TF: #1}}
\newcommand{\nsg}[1]{\textcolor{cyan}{#1}}
\newcommand{\saeed}[1]{\textcolor{blue}{SF: #1}}
\newcommand{\fdg}[1]{\textcolor{green}{FDG: #1}}
\newcommand{\jc}[1]{\textcolor{magenta}{JC: #1}}

\def\CovDev{D}
\def\Res{{\mathcal R}}
\def\Gammaflat{\hat \Gamma}
\def\metricflat{\hat \gamma}
\def\Dflat{\hat {\mathcal D}}
\def\part_n{\partial_\perp}

\def\Lie{\mathcal{L}}
\def\A{\mathcal{X}}
\def\Aphi{\A_{\phi}}
\def\hAphi{\hat{\A}_{\phi}}
\def\E{\mathcal{E}}
\def\Ham{\mathcal{H}}
\def\M{\mathcal{M}}
\def\R{\mathcal{R}}
\def\p{\partial}

\def\hg{\hat{\gamma}}
\def\hA{\hat{A}}
\def\hD{\hat{D}}
\def\hE{\hat{E}}
\def\hR{\hat{R}}
\def\hcA{\hat{\mathcal{A}}}
\def\hDelt{\hat{\triangle}}

\def\na{\nabla}
\def\dif{{\rm{d}}}
\def\non{\nonumber}
\newcommand{\erf}{\textrm{erf}}

\renewcommand{\t}{\times}

\long\def\symbolfootnote[#1]#2{\begingroup%
\def\thefootnote{\fnsymbol{footnote}}\footnote[#1]{#2}\endgroup}


\title{Dynamical formation and stability of fermion-boson stars}
 
\author{Fabrizio Di Giovanni}
\affiliation{Departamento de
  Astronom\'{\i}a y Astrof\'{\i}sica, Universitat de Val\`encia,
  Dr. Moliner 50, 46100, Burjassot (Val\`encia), Spain}
  
\author{Saeed Fakhry}
\affiliation{Department of Physics, Shahid Beheshti University, G. C., Evin, Tehran 19839, Iran}
\affiliation{Departamento de Astronom\'{\i}a y Astrof\'{\i}sica, Universitat de Val\`encia, Dr. Moliner 50, 46100, Burjassot (Val\`encia), Spain}

\author{Nicolas Sanchis-Gual}
\affiliation{Centro de Astrof\'\i sica e Gravita\c c\~ao - CENTRA, Departamento de F\'\i sica,
Instituto Superior T\'ecnico - IST, Universidade de Lisboa - UL, Avenida
Rovisco Pais 1, 1049-001, Portugal}

\author{Juan Carlos Degollado} 
\affiliation{Instituto de Ciencias F\'isicas, Universidad Nacional Aut\'onoma 
de M\'exico, Apdo. Postal 48-3, 62251, Cuernavaca, Morelos, M\'exico}

\author{Jos\'e A. Font}
\affiliation{Departamento de
  Astronom\'{\i}a y Astrof\'{\i}sica, Universitat de Val\`encia,
  Dr. Moliner 50, 46100, Burjassot (Val\`encia), Spain}
\affiliation{Observatori Astron\`omic, Universitat de Val\`encia, C/ Catedr\'atico 
  Jos\'e Beltr\'an 2, 46980, Paterna (Val\`encia), Spain}




\begin{abstract} 
Gravitationally bound structures composed by fermions and scalar particles 
known as fermion-boson stars are regular and static configurations obtained 
by solving the coupled Einstein-Klein-Gordon-Euler (EKGE) system. In this work, we 
discuss one possible scenario through which these fermion-boson stars may 
form by solving numerically the EKGE system under the simplifying assumption of 
spherical symmetry. Our initial configurations assume an already existing neutron 
star surrounded by an accreting cloud of a massive and complex scalar field. The results of our 
simulations show that once part of the initial scalar field is expelled via gravitational 
cooling the system gradually oscillates around an equilibrium configuration that is asymptotically 
consistent with a static solution of the system. The formation 
of fermion-boson stars for large positive values of the coupling constant in the 
 self-interaction term of the scalar-field potential reveal the presence of 
a node in the scalar field. This suggests that a fermionic core may 
help stabilize configurations with nodes in the bosonic sector, as happens for 
purely boson stars in which the ground state and the first excited state coexist.
\end{abstract}




\maketitle

\vspace{0.8cm}

\section{Introduction}

Identifying the relevance scalar fields may have for astrophysics and cosmology, in particular as potential components of the dark matter content of the universe, has long received considerable attention~\cite{Weinberg:1977ma, Preskill:1982cy, Matos:2000ss,Matos:2000ng}. Different scalar fields have been considered, namely the dilaton in string theories~\cite{Gasperini:1994xg, Svrcek:2006yi}, the Higgs boson in the standard model of particle physics~\cite{Higgs:1964ia, ATLAS:2012oga}, the inflaton in studies of the early universe~\cite{Guth:1980zm, Langlois:2004de}, or the axion as a  possible component of cold dark matter~\cite{Kawasaki:2013ae,  Arvanitaki:2009fg, Hui:2016ltb, Klaer:2017ond}.
 
It has been argued that ultralight bosons form localized and coherently oscillating configurations
very similar to Bose-Einstein condensates~\cite{Sin:1992bg, Chavanis:2011cz}. When the mass of the bosonic particle is around $10^{-22}$ eV~\cite{Matos:1999et,Hu:2000ke} these condensates provide an alternative to the standard approach to explain large-scale structure formation through dark-matter seeds. For heavier bosons, the bound configurations are smaller and may have the typical size and mass of a sellar compact object such as a neutron star. These objects are generically known as boson stars~\cite{Jetzer:1991jr}.

Boson stars are gravitationally bound configurations of scalar particles. 
Since the seminal works of Kaup~\cite{Kaup:1968zz} and Ruffini and Bonnazola~\cite{Ruffini:1969qy} their description has been generalised in several ways including self-interaction~\cite{Colpi:1986ye}, charge \cite{Jetzer:1989av}, rotation~\cite{Yoshida:1997qf, Schunck:1996he}, oscillating soliton stars~\cite{Seidel91}, stars with more than a single scalar field~\cite{Alcubierre:2018ahf,Jaramillo:2020rsv}, and even vector fields (in which case the bosonic star is known as a Proca star~\cite{brito2016proca}). Reviews on the subject can be found in references~\cite{Schunck:2003kk, liebling2017dynamical}.

If such bosonic configurations could form from some primordial gas, it is natural to assume that other particles, such as fermions, could also be present during the condensation. Therefore, it would seem theoretically possible that objects made out of a mixture of both bosons and fermions might also form. Even if the original configurations were mainly composed by either bosons or fermions, they could be susceptible to further capture fermions and bosons through accretion giving rise to mixed configurations. It is thus a theoretically interesting question to investigate the properties of these macroscopic composites of fermions and bosons, referred in the literature as fermion-boson stars~\cite{HENRIQUES1990511,valdez2013dynamical,brito2015accretion,brito2016interaction,valdez2020fermion} 
and to discuss possible means by which they might form. This is the focus of this paper. Here we propose a dynamical scenario in which a fermionic star (modelled as a polytropic star for simplicity) accretes part of the scalar field, while part of it is radiated to infinity, and a mixed fermion-boson star forms.

The gravitational condensation of a primordial gas and the subsequent radiation of part of the bosonic field has been dubbed  gravitational cooling and has been addressed in~\cite{Seidel:1993zk} for purely scalar fields and in~\cite{di2018dynamical} for vector fields. Using numerical-relativity simulations those studies have shown the dynamical formation of boson stars and Proca stars, respectively, under the assumption of spherical symmetry. In order to be astrophysically relevant, a gravitationally bound system that forms dynamically must be stable for times much longer than its characteristic dynamical timescale. The stability properties of boson stars  have been considered in~\cite{1989NuPhB.315..477L, Hawley:2000dt, Gleiser:1988rq, Gleiser:1989a, Balakrishna:1997ej, Guzman09, sanchis2017numerical,sanchis2019nonlinear}. In Ref.~\cite{Seidel90} Seidel and Suen discussed the dynamical evolution of perturbed boson stars finding, in particular, that unstable stars migrate to the stability region of static configurations which suggests the formation of boson stars under generic initial conditions. Further studies on the formation of boson stars were performed in~\cite{Seidel:1993zk} in general relativity and in~\cite{Guzman:2004wj, Guzman:2006yc} in the Newtonian regime. These studies concluded that self-gravitating, scalar-field stellar systems settle down into equilibrium configurations. We note that this conclusion does not only apply to the scalar case but it is also valid for the vector counterparts of boson stars, i.e.~Proca stars, as has recently been reported in~\cite{di2018dynamical}.

The purpose of this work is twofold: on the one hand we aim to describe the dynamical formation of fermion-boson stars; on the other hand, we will analyse the stability properties of those configurations considering a strong self-interaction term in the Klein-Gordon potential of the bosonic part. For this study, and for the sake of simplicity, we shall focus on fermion-boson stars assuming spherical symmetry. The starting point of our analysis assumes a preexisting neutron star (described with a polytropic equation of state) surrounded by a cloud of scalar field. Different initial configurations are evolved in time using numerical-relativity simulations. We find that the fermionic star is able to capture part of the scalar field and the new system evolves toward an almost static configuration giving rise to a stable fermion-boson star. In addition to show that the dynamical formation of mixed stars is possible we also obtain  the corresponding equilibrium configurations for fermion-boson stars with different values of the self-interaction potential and we study their stability properties under spherical perturbations.

This paper is organized as follows: in Section~\ref{sec2} we introduce the matter model we employ to describe fermion-boson stars and set up the basic equations. Section~\ref{sec:ini} addresses the initial data for the dynamical formation of the mixed stars and the initial static configurations considering a self-interaction term in the bosonic sector. The numerical framework for our simulations is described in Section~\ref{sec:numerics} while in Section~\ref{results} the results of the evolutions are presented. Finally, our conclusions and final remarks are reported in Section ~\ref{sec:conclusions}. Our units are such that the relevant fundamental constants are equal to one $(G=c=\hbar=1)$.

\section{Setup}
\label{sec2}

In this study we consider that bosonic and fermionic matter only interact  through gravity. Therefore, our model is described by a total stress-energy tensor which is the sum of two contributions, one from a perfect fluid and one from a complex scalar field:
\begin{eqnarray}
T_{\mu\nu}&=& T_{\mu\nu}^{\rm{fluid}} + T_{\mu\nu}^{\phi} ,
\end{eqnarray}
where
\begin{eqnarray}
T_{\mu\nu}^{\rm{fluid}}&=& [\rho(1+\epsilon) + P] u_{\mu}u_{\nu} + P g_{\mu\nu}, \\
T_{\mu\nu}^{\phi}&=& - \frac{1}{2}g_{\mu\nu}\partial_{\alpha}\bar{\phi}\partial^{\alpha}\phi - V(\phi) \nonumber \\
 &+& \frac{1}{2}(\partial_{\mu}\bar{\phi}\partial_{\nu}\phi+\partial_{\mu}\phi\partial_{\nu}\bar{\phi}) .
\end{eqnarray}

The perfect fluid is described by its pressure $P$, its rest-mass density $\rho$, and its internal energy $\epsilon$, while $u^{\mu}$ is the fluid 4-velocity. We consider a quartic self-interaction potential for the scalar field $\phi$
\begin{equation} \label{s0}
V(\phi) =  \frac{1}{2} \mu^2\bar{\phi}\phi+\frac{1}{4}\lambda(\bar{\phi}\phi)^2,
\end{equation}
where $\mu$ is the mass of the bosonic particle and $\lambda$ is the self-interaction parameter; the bar symbol  in the last two equations denotes complex conjugation. The equations of motion are given by the conservation laws of the stress-energy tensor and the baryonic particles
\begin{eqnarray} 
\nabla_{\mu}T^{\mu\nu}_{\rm{fluid}} = 0, \label{conservation_laws1} \\
\nabla_{\mu}(\rho u^{\mu}) = 0, \label{conservation_laws2} 
\end{eqnarray}
for the fermionic matter, and by the Klein-Gordon equation
\begin{equation}
\nabla_{\mu}\nabla^{\mu} \phi= \mu^2 \phi + \lambda |\phi|^2 \phi \label{Klein-Gordon}
\end{equation}
for the complex scalar field, together with the Einstein equation $G_{\mu\nu}=8\pi T_{\mu\nu}$ governing the spacetime dynamics. Differential operator $\nabla_{\mu}$ is the covariant derivative with respect to the 4-metric $g_{\mu\nu}$. The set of equations~\eqref{conservation_laws1}-\eqref{conservation_laws2} is closed by an equation of state (EoS) for the fluid. We consider both the polytropic EoS and the ideal-gas EoS, 
\begin{equation}\label{EOS}
P= K \rho^{\Gamma} = (\Gamma-1)\rho\epsilon\,.
\end{equation}
The polytropic EoS is employed to build the equilibrium initial data while  the $\Gamma$-law is used for the evolutions as it would allow to take into account eventual shock-heating (thermal) effects. All equilibrium models we consider are constructed using $K=100$ and $\Gamma = 2$. In the next subsections we specify our choice for the metric and the relevant equations for both the construction of the static models and the evolution.

\subsection{Basic equations for the equilibrium configurations} \label{stat}

Our formalism for the construction of  equilibrium configurations of fermion-bosn stars relies on the choice of a spherically symmetric metric in Schwarzschild coordinates
\begin{equation} \label{Schwarzschild_metric}
ds^2 = -\alpha(r)^2 dt^2 + \tilde{a}(r)^2 dr^2 + r^2 ( d\theta^2 + \sin{\theta}^2 d\varphi^2),
\end{equation}
written in terms of two geometrical functions $\tilde{a}(r)$ and $\alpha(r)$. We set a harmonic time dependence ansatz for the complex scalar field $\phi(t, r) = \phi(r) e^{-i\omega t}$ where $\omega$ is its eigenfrequency, and we consider the quartic self-interaction potential for the field given by Eq.~\eqref{s0}. We replace the self-interaction parameter $\lambda$ by the dimensionless variable $\Lambda$, defined as
\begin{align}
\Lambda=\frac{M_{\rm Pl}^{2}\lambda}{4\pi \mu^{2}},
\end{align}
in which $M_{\rm Pl} = \sqrt{\frac{\hbar c}{G}}$ indicates the Planck mass (which is one in our units).
 In the following we consider a scaled radial coordinate $r\rightarrow r\mu$ (together with $M\rightarrow M\mu$, $t\rightarrow t\mu$, $\omega\rightarrow \omega/\mu$).
Assuming a static fluid, $u^{\mu}=(-1/\alpha,0,0,0)$, Einstein's equations lead to the following ordinary differential equations (ODEs)
\begin{align} \label{s1}
\frac{d\tilde{a}}{dr} & = \frac{\tilde{a}}{2}\left(\frac{1-\tilde{a}^{2}}{r} +4\pi r \biggl[\biggl(\frac{\omega^{2}}{\alpha^{2}}+\mu^{2}+\frac{\lambda}{2}\phi^{2}\biggl) \tilde{a}^{2}\phi^{2}\right.\nonumber \\
		& \left. +\Psi^{2}+2\tilde{a}^{2}\rho(1+\epsilon)\biggl]\right.\biggl),
\end{align}
\begin{align} \label{s2}
\frac{d\alpha}{dr} & = \frac{\alpha}{2}\left(\frac{\tilde{a}^{2}-1}{r} +4\pi r \biggl[\biggl(\frac{\omega^{2}}{\alpha^{2}}-\mu^{2}-\frac{\lambda}{2}\phi^{2}\biggl) \tilde{a}^{2}\phi^{2}\right.\nonumber \\
		& \left.+\Psi^{2}+2\tilde{a}^{2}P\biggl] \right.\biggl),
\end{align}
\begin{align} \label{s3}
\frac{d\phi}{dr} = \Psi ,
\end{align}
\begin{align} \label{s4}
\frac{d\Psi}{dr} & = -\left(1+\tilde{a}^{2}-4\pi r^{2}\tilde{a}^{2} (\mu^{2}\phi^{2}+\frac{\lambda}{2}\phi^{4}\right.\nonumber \\
	&+\rho(1+\epsilon)-P)\biggl)\frac{\Psi}{r}-\left(\frac{\omega^{2}}{\alpha^{2}}-\mu^{2}-\lambda\phi^{2}\right)\tilde{a}^{2}\phi^{2},
\end{align}
\begin{align} \label{s5}
\frac{dP}{dr} = -[\rho(1+\epsilon)+P]\frac{\alpha^{\prime}}{\alpha},
\end{align}
where the prime indicates the derivative with respect to $r$.  The system is closed by the EoS \eqref{EOS}. To solve these equations it is necessary to apply certain initial and boundary conditions that are consistent with the geometry and physical behavior of the mixed stars. In Section \ref{sec:initial_static} we will introduce these conditions.

\subsection{Basic equations for the evolution} \label{sec:basic}

For the numerical evolutions we consider a spherically symmetric metric in isotropic coordinates
\begin{equation}\label{isotropic_metric}
ds^2 = -\alpha(\hat{r})^2 dt^2 + \psi(\hat{r})^4 \gamma_{ij} (dx^{i} + \beta^{i}dt)(dx^{j} + \beta^{j}dt),
\end{equation} 
where $\alpha$ is the lapse function and $\beta^{i}$ is the shift vector. The spatial 3-dimensional metric components are
\begin{eqnarray}
\gamma_{ij} dx^idx^j = a(\hat{r})d\hat{r}^2 + b(\hat{r})\hat{r}^2 (d\theta^2 + \sin{\theta}^2 d\varphi^2) \,.
\end{eqnarray}
We note that $a$ and $\tilde{a}$ should not be confused as they are different functions; $a(\hat{r})$ and $b(\hat{r})$ are the metric functions for the isotropic metric, $\hat{r}$ denotes the isotropic radial coordinate (see section \ref{sec:evolutions} for details) and $\psi^4 \equiv e^{4\chi}$ is the conformal factor.
To simplify the notation we will substitute $\hat{r} \rightarrow r$ in the following, keeping in mind that all equations and definitions refer nonetheless to the isotropic radial coordinate.

Our choice of evolution equations for the spacetime variables follows Brown's covariant form~\cite{Brown:2009,Alcubierre:2010is} of the Baumgarte-Shapiro-Shibata-Nakamura (BSSN) formulation of Einstein's equations~\cite{Nakamura87,Shibata95, Baumgarte98}.
The evolved quantities used in this work are the spatial metric $\gamma_{ij}$, the conformal factor $\chi$, the trace of the extrinsic curvature $K$, its traceless part $A_a = A^r_r$, $A_{b}=A^{\theta}_{\theta}=A^{\varphi}_{\varphi}$, and the radial component of the so-called conformal connection functions $\Delta^r$ (see~\cite{Shibata95, Baumgarte98} for definitions). 

We will not report here explicitly the full system of evolution equations as it can be found e.g.~in Ref.~\cite{Montero:2012yr}. We remind the reader that the equations involve matter source terms arising from suitable projections of the total stress-energy tensor $T_{\mu\nu}$, namely the energy density $\mathcal{E}$, the momentum density $j_{i}$ measured by a normal observer $n^{\mu}$, and the spatial projection of the energy-momentum tensor $S_{ij}$. These quantities read as
\begin{align}
\mathcal{E}&= n^{\mu}n^{\nu}T_{\mu \nu}, \\
j_i&=-\gamma_{i}^{\mu}n^{\nu}T_{\mu \nu}, \\
S_{ij}&= \gamma_{i}^{\mu} \gamma_{j}^{\nu} T_{\mu \nu}.
\end{align}
In our setup these quantities are obtained by adding up the contributions of both the fluid and the scalar field. The explicit expressions we use are listed at the end of this section.

The gauge conditions we employ 
in our simulations are the so-called ``non-advective $1+$log'' gauge condition for the lapse function $\alpha$ and a variation of the Gamma-driver condition for the shift vector $\beta^r$. Further details regarding the BSSN evolution equations, gauge conditions, and the formalism for the hydrodynamic equations can be found in~\cite{Montero:2012yr}. 

Following our previous work~\cite{Sanchis-Gual:2015bh} we use two auxiliary variables
\begin{eqnarray}
\Pi &=& \frac{1}{\alpha}(\partial_{t} - \beta^r\partial_{r})\phi, \label{Pi_scalar} \\
\Psi &=& \partial_r \phi, \label{Psi_scalar}
\end{eqnarray}
to cast the Klein-Gordon equation \eqref{Klein-Gordon} as a first-order system of evolution equations:
\begin{align}
\partial_t \phi & = \beta^r\partial_r \phi + \alpha \Pi, \\
\partial_t \Pi & = \beta^r \partial_r \Pi + \frac{\alpha}{a e^{4\chi}} \biggl[\partial_r\Psi +  \Psi \biggl(\frac{2}{r} - \frac{\partial_r a}{2a} + \frac{\partial_r b}{b} \nonumber\\
         &+ 2 \partial_r{\chi}\biggr)\biggr] + \frac{\Psi}{a e^{4\chi}} + \alpha K \Pi - \alpha(\mu^2 + \lambda \phi \bar{\phi})\phi, \\
\partial_t \Psi & = \beta^r\partial_r \Psi + \Psi \partial_r \beta^r + \partial_r(\alpha \Pi).
\end{align}

Finally, the system of equations is closed by two constraint equations, namely the Hamiltonian constraint and the momentum constraint, which read as
\begin{align}
\mathcal{H} & = R - (A_{a}^2 + 2 A_{b}^2) + \frac{2}{3} K^2 - 16\pi \mathcal{E} = 0, \label{Hamiltonian constraint} \\
\mathcal{M}_{r} & = \partial_{r}A_{a} - \frac{2}{3} \partial_{r}K + 6A_{a}\partial_{r}\chi + \nonumber \\
		& (A_{a}-A_{b}) (\frac{2}{r} +\frac{\partial_{r}b}{b}) - 8\pi j_{r} = 0, \label{Momentum constraint}
\end{align} 
where $R$ is the Ricci scalar. 
 
The bosonic contribution to the matter source terms are
\begin{align}
\mathcal{E}^{\phi}&= \frac{1}{2} \left( \bar{\Pi}\,\Pi + \frac{\bar{\Psi}\Psi}{e^{4\chi}a}\right) + \frac{1}{2} \mu^2 \bar{\phi}\phi + \frac{1}{4} \lambda (\bar{\phi}\phi)^2 \label{rhomat_phi}\\
j_r^{\phi}&= - \frac{1}{2} (\bar{\Pi}\Psi + \bar{\Psi}\Pi), \label{j_phi}\\
S_{a}^{\phi}&= \frac{1}{2} \left(\bar{\Pi}\,\Pi + \frac{\bar{\Psi}\Psi}{e^{4\chi}a}\right) - \frac{1}{2} \mu^2 \bar{\phi}\phi - \frac{1}{4} \lambda (\bar{\phi}\phi)^2 \\
S_{b}^{\phi}&= \frac{1}{2} \left(\bar{\Pi}\,\Pi - \frac{\bar{\Psi}\Psi}{e^{4\chi}a}\right) - \frac{1}{2} \mu^2 \bar{\phi}\phi - \frac{1}{4} \lambda (\bar{\phi}\phi)^2
\end{align}
where $S_a = S^r_r$ and $S_b=S^{\theta}_{\theta} = S^{\varphi}_{\varphi}$. Correspondingly, the fermionic contribution to those source terms read 
\begin{align}
\mathcal{E}^{\rm{fluid}}&= \left[\rho\,(1 + \epsilon) + P \right] W^2 - P, \\
j_r^{\rm{fluid}}&= e^{4\chi} a \left[\rho\,(1+\epsilon) + P\right] W^2 v^{r}, \\
S_{a}^{\rm{fluid}}&=  e^{4\chi} a \left [\rho\,(1+\epsilon) + P\right]W^2  v^{r} + P, \\
S_{b}^{\rm{fluid}}&= P,
\end{align}
where $W = \alpha u^{t}$ is the Lorentz factor and $v^r$ is the radial component of the fluid 3-velocity. 

\section{Initial Data} \label{sec:ini}

As mentioned in the introduction we consider two different physical situations in this paper, namely the dynamical formation of a fermion-boson star and the stability properties of different equilibrium models of such stars. In the following we discuss the corresponding initial data for either situation.

\subsection{Dynamical formation} \label{sec:initial_formation}

To study the dynamical formation of a mixed star we begin with a stable fermionic star (FS) model surrounded by a dilute cloud of bosonic particles. This cloud accretes on to the FS under the gravitational pull of the latter. Suitable initial data describing this system are secured after solving the Hamiltonian constraint~\eqref{Hamiltonian constraint} and the momentum constraint~\eqref{Momentum constraint}. To do so we assume a harmonic time dependence for the scalar field and choose a Gaussian radial distribution for the cloud, yielding 
\begin{equation}
\phi(r,t) = A_{0}\,e^{-\frac{r^2}{\sigma^2}} e^{-i\omega t}, \label{scalarfield}
\end{equation}
where parameters $A_{0}$ and $\sigma$ are the amplitude and the width of the Gaussian profile, respectively, and $\omega$ is the initial frequency of the field.

To solve the constraints we initially consider the spacetime of an isolated spherically symmetric FS by solving the Tolman-Oppenheimer-Volkoff equation. Next, we add to this solution the dilute cloud of bosonic matter described by~\eqref{scalarfield}. The time symmetry condition, $K_{ij}=0$, and the conformally flat condition,  $a=b=1$, yield the following initial values for a set of spacetime variables
\begin{equation}\label{parameters}
\begin{split}
\beta^r&=0,  \\
K&=0, \\
A_a&=A_b=0, \\
\Delta^r&=0. \\
\end{split}
\end{equation}
while the values of the conformal factor $\psi$ and of the lapse function $\alpha$ are inferred directly from the FS spacetime. Starting with these initial conditions we solve numerically the Hamiltonian constraint~\eqref{Hamiltonian constraint} using the procedure described in~\cite{Sanchis-Gual:2015bh}. This yields an updated value of the conformal factor $\psi$ and of the $\gamma_{rr}$ metric component.

Due to the harmonic time dependence of the scalar field, it follows that $j_{r}^{\phi}$ defined by~\eqref{j_phi} is zero.  This means that the scalar field does not contribute to the momentum constraint equation~\eqref{Momentum constraint}. Therefore, considering~\eqref{parameters} the momentum constraint is analytically solved.

\subsection{Equilibrium configurations} \label{sec:initial_static}
In Section~\ref{stat} we introduced the basic equations to construct the static models of mixed stars. To solve the set of equations, namely equations~\eqref{s1}-\eqref{s5} and the EoS \eqref{EOS}, we need to construct suitable initial data which are compatible with the physical and geometrical conditions of the stellar configurations. The system of ODEs becomes an eigenvalue problem for the frequency $\omega$, which is a function of two parameters, the central value of the scalar field, $\phi_{c}$, and of the fermionic density, $\rho_c$. We make use of the two-parameter shooting method to find the solution for $\omega$. Once this is found and the central values of all variables are available, we use a 4th-order Runge-Kutta method to solve the ODEs and reconstruct the radial profiles of the solution. 

We require the condition of regularity at the origin to be satisfied for the metric functions. At the outer boundary we employ the values provided by the Schwarzschild solution at the outer radius, which do not depart much from the values of a flat metric, together with a vanishing scalar field value. Hence, the boundary conditions for solving the set of ODEs can be defined as follows
\begin{eqnarray}
&\tilde{a}(0) = 1, \hspace{0.3cm} & \phi(0) = \phi_{c}, \nonumber\\
&\alpha(0) = 1, \hspace{0.3cm} &  \lim_{r\rightarrow\infty}\alpha(r)=\lim_{r\rightarrow\infty}\frac{1}{\tilde{a}(r)},\nonumber\\
& \Psi(0)=0, \hspace{0.3cm} & \lim_{r\rightarrow\infty}\phi(r)=0, \nonumber\\
&\rho(0) = \rho_{c},  & \hspace{0.3cm}  P(0)=K\rho_{c}^{\Gamma},  \hspace{0.3cm} \lim_{r\rightarrow\infty}P(r)=0.
\end{eqnarray}

Once the solution is found, one can define the total gravitational mass based on the value of the metric coefficients at infinity
\begin{align} \label{mass}
M_T=\lim_{r\longrightarrow\infty}\frac{r}{2}\left(1-\frac{1}{\tilde{a}^2}\right),
\end{align}
which coincides with the Anowitt-Desser-Misner (ADM) mass at infinity. As the Klein-Gordon Lagrangian for a complex scalar field exhibits invariance under global U(1) transformations $\phi \rightarrow \phi\,e^{i\delta}$, Noether's theorem predicts the existence of a conserved charge which can be associated with the number of bosonic particles $N_B$; moreover, the conservation of the baryonic number provides a definition of the number of fermionic particles $N_F$. These two quantities can be evaluated by integrating their volume density as follows
\begin{align}
N_B = 4 \pi \int \frac{\tilde{a} \omega \phi^2 r^2}{\alpha} dr, \hspace*{0.5cm} N_F= 4 \pi \int \tilde{a} \rho r^2 dr.
\end{align} 
These quantities will be used to determine the conservation of the number of particles, both bosons and fermions during the numerical evolutions. Finally, we evaluate the radius of the bosonic (fermionic) contribution to the mixed star, $R_{B}$($R_{F}$), as the radius of the sphere containing $99\%$ of the corresponding particles.

\begin{figure}[t]
\begin{minipage}{1\linewidth}
\includegraphics[width=1.0\textwidth]{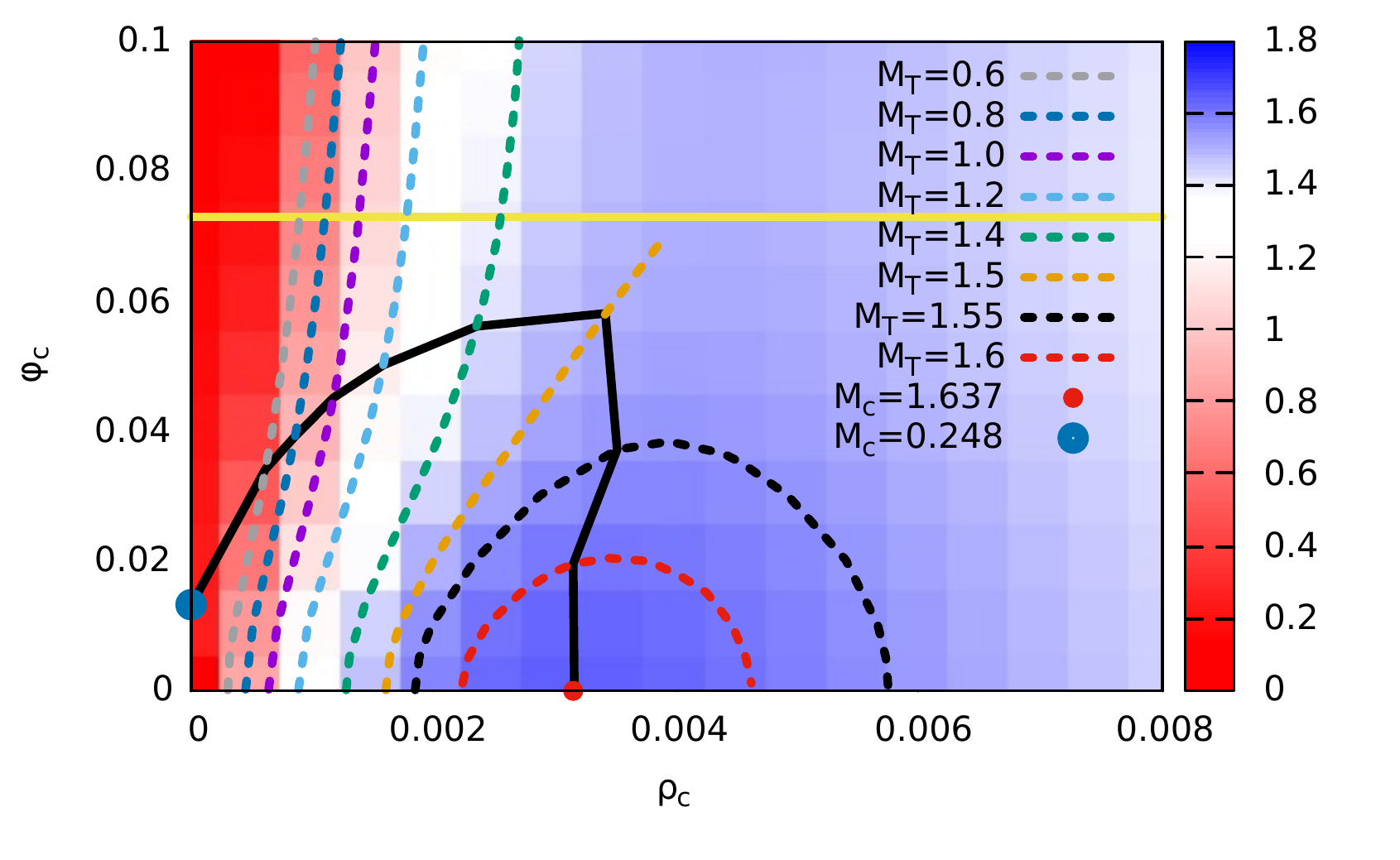}  
\includegraphics[width=1.0\textwidth]{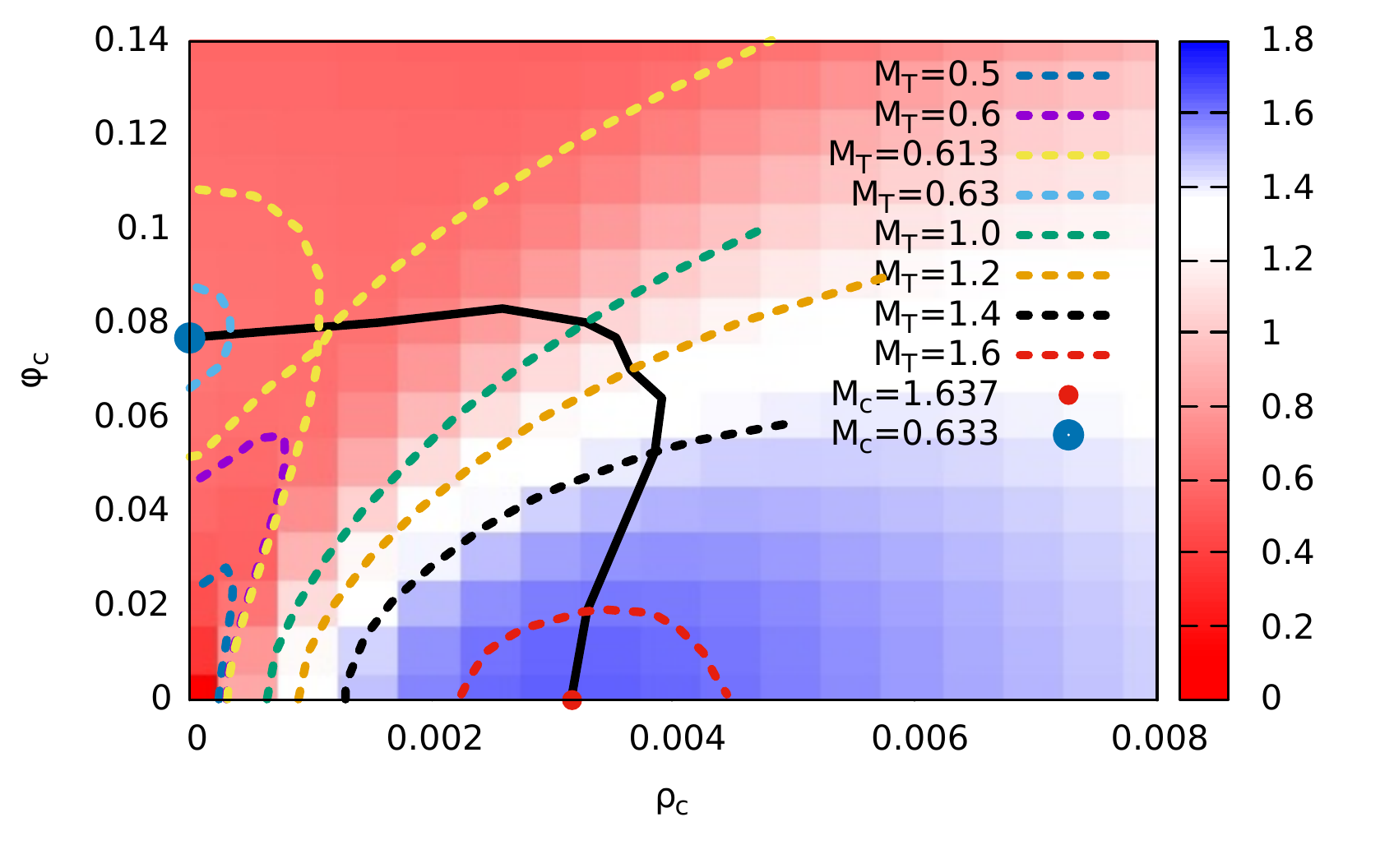} 
\includegraphics[width=1.0\textwidth]{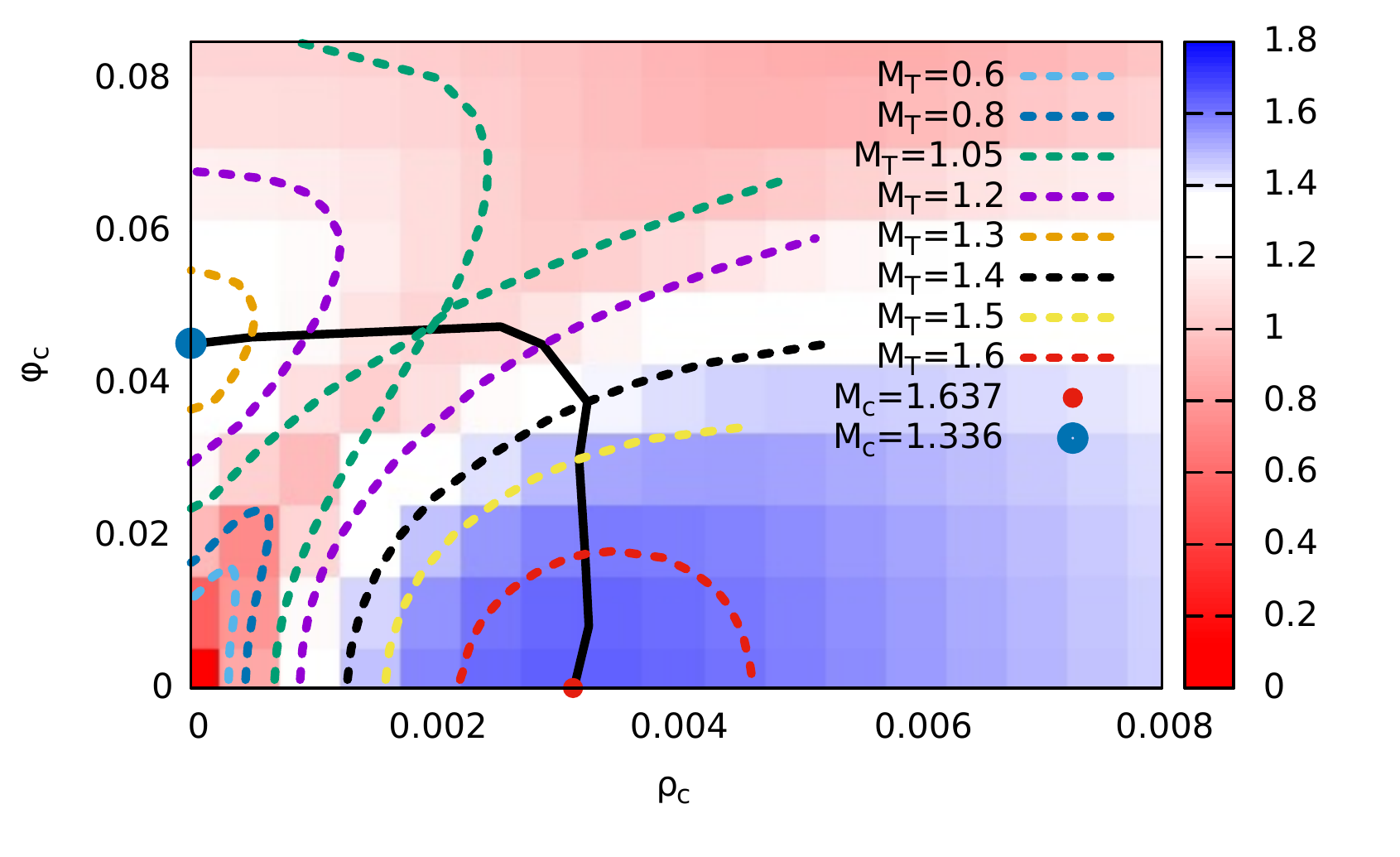} 
\caption{Equilibrium configurations of fermion-boson stars for $\Lambda = -30$ (top), $\Lambda = 0$ (middle), and $\Lambda = 30$ (bottom). Dashed lines correspond to models with the same total mass $M_{T}$. The black solid line depicts the boundary between stable and unstable models, and the solid yellow line for the case $\Lambda = -30$ indicates the maximum value of $\phi_c$ that assures the non-negativity of the scalar field potential $V(\phi)$ in the entire spatial domain. } 
\label{fig:existence_plot} 
\end{minipage}
\end{figure}

As mentioned before, the construction of the static solutions for the fermion-boson stars depends on two parameters, namely the central fluid density $\rho_c$ and the central value of the scalar field $\phi_c$. We can therefore express the mass of the system~\eqref{mass} as a function of these two parameters $M_{T}(\rho_c, \phi_c)$ as we depict in figure \ref{fig:existence_plot} for three different values of $\Lambda$. In the case of non-rotating boson stars the parameter space is 1-dimensional and stability theorems~\cite{Lee:1988av} indicate that for each value of $\Lambda$ there exists a critical mass such that $dM_{T}/d\phi_c=0$. These critical points indicate the transitions between the stability and the instability region of the parameter space. Analogous transitions in stability occur in fermionic stars (see e.g.~\cite{Cook1994,Friedman1988}). In the case of fermion-boson stars, as the parameter space is 2-dimensional, the analysis is more involved. Following \cite{HENRIQUES1990511} we define the critical points as the values of the pair ($\rho_c$,$\phi_c$) such that the conditions
\begin{align}
\frac{\partial N_{B}}{\partial \rho_c} \Bigr|_{\substack{M=\rm{constant}}} = \frac{\partial N_{F}}{\partial \rho_c}\Bigr|_{\substack{M=\rm{constant}}} = 0, \nonumber \\
\frac{\partial N_{B}}{\partial \phi_c}\Bigr|_{\substack{M=\rm{constant}}} = \frac{\partial N_{F}}{\partial \phi_c}\Bigr|_{\substack{M=\rm{constant}}} = 0,
\end{align}
are satisfied. In Fig.~\ref{fig:existence_plot} we show several curves of constant mass in the parameter space (dashed colored lines). For each point of the curves we evaluate the number of bosons $N_{B}$ and fermions $N_{F}$. If we start from a purely FS configuration (a point on the horizontal axis in Fig.~\ref{fig:existence_plot}) and we move along a curve of fixed mass changing the values of $\phi_c$ and $\rho_c$, the number of bosons increases and the number of fermions deacreases up to a critical point in the parameter space where a maximum is found for $N_{B}$ and a minimum for $N_{F}$. If we start from a pure boson star (a point on the vertical axis), the behavior is the opposite, with $N_{B}$ decreasing up to a minimum and $N_{F}$ increasing up to a maximum. For each value of the mass, these critical points signal the boundary between the stability and instability regions. The black solid line in Fig.~\ref{fig:existence_plot}  represents these boundaries in the parameter space for the values of $\Lambda=\{-30,0,30\}$. This construction follows the same approach laid out in~\cite{valdez2013dynamical,valdez2020fermion}.

As FS do not depend on $\Lambda$ their threshold mass  is constant for all values of $\Lambda$ and equal to $M_c=1.637$. On the contrary, for boson stars the threshold mass changes with $\Lambda$. In particular the threshold masses for our pure boson star models are $M_c= 0.248$, $0.633$, and $1.336$, for $\Lambda = -30$, $0$, and $30$, respectively. For fermion-boson stars, one can observe that for the same point in the parameter space with fixed values of $\phi_c$ and $\rho_c$, the total mass decreases (increases) for positive (negative) values of $\Lambda$, with respect to the $\Lambda = 0$ case.

We point out that considering negative values of $\lambda$ raises the issue that the scalar potential $V(|\phi|) = \frac{1}{2}\mu^2 |\phi|^2 + \frac{1}{4} \lambda (|\phi|^2)^2$ is not bounded from below and can become negative, breaking the weak-energy condition (see e.g.~the discussion in~\cite{Barcelo2000}). For $\Lambda =-30$  we evaluate the maximum central value of $\phi$ that ensures the non-negativity of the scalar field potential, yielding $\phi_{c}=0.0728$. We depict in the top plot of Fig.~\ref{fig:existence_plot} a horizontal yellow line at this value. We disregard all stellar models above this line as they may give rise to naked singularities.

\section{Numerical framework} 
\label{sec:numerics}

The numerical evolutions of the Einstein-Klein-Gordon-Euler system are performed with the numerical-relativity code originally developed by~\cite{Montero:2012yr} and upgraded to take into account the complex scalar-field equations in~\cite{escorihuela2017quasistationary}. This computational infrastructure has been extensively used by our group in studies of fundamental bosonic fields in strong-gravity spacetimes (see e.g.~\cite{Sanchis-Gual:2015bh,Sanchis-Gual:2015sms,Sanchis-Gual:2015lje,sanchis2017numerical,di2018dynamical}). 

The time update of the evolution equations is evaluated using a Partially Implicit Runge-Kutta method developed by~\cite{Isabel:2012arx, Casas:2014}. In this scheme the operators in the right-hand-side of the BSSN evolution equations are divided into operators which are evaluated explicitly, and operators carrying geometrical singularities which are evolved implicitly using the updated values of the first ones. This allows to handle potential numerical instabilities arising from $1/r$ terms in the equations. While the construction of the equilibrium configurations employs Schwarzschild coordinates and an equally spaced linear grid, the dynamical evolutions  make use of isotropic coordinates and a logarithmic grid. More precisely, the computational domain of the simulations is covered with an isotropic grid which is composed by two different patches, a geometrical progression up to a certain radius and an hyperbolic cosine in the exterior part. This  allows  to place the outer boundary sufficiently far from the origin and prevent the effects of reflections.  Further details about the computational grid can be found in~\cite{Sanchis-Gual:2015sms}. The minimum resolution we employ in our simulations is $\Delta r=0.0125$. The inner boundary is then set at $r_{\rm{min}}=\Delta r/2$ and the outer boundary is at $r_{\rm{max}}=6000$. The time step is given by $\Delta t=0.3\,\Delta r$ in order to obtain long-term stable simulations. We add 4th-order Kreiss-Oliger numerical dissipation terms to the evolution equations to damp out spurious, high-frequency numerical noise. All advection terms (such as $\beta^{r}\partial_{r}f$) are treated with an upwind scheme. At the outer boundary we impose radiative boundary conditions.

\section{Results}
\label{results}

\subsection{Dynamical formation of fermion-boson stars}

\begin{table*}[t]
\caption{Initial models leading to stable fermion-boson stars. The vertical lines divide the initial parameters (left), from the physical quantities evaluated at the end of the time evolution (center) and from the physical quantities of the corresponding equilibrium configuration (right). Note that as model MS5 forms an excited state, we cannot compare it with a nodeless static solution. Columns on the left indicate the central rest-mass density $\rho_{c}$, the self-interaction parameter $\Lambda$, and the amplitude of the scalar field profile $A_{0}$. Columns at the center indicate the scalar field frequencies $\omega_{n}$, the fermionic energy $E^{\rm{fluid}}_{30}$ contained in a sphere of radius $r=30$, the bosonic energy $E^{\phi}_{30}$, and the ratio between number of bosons and fermions $N^{B}_{30}/N^{F}_{30}$. Columns on the right side indicate the frequency $\omega$, the fermionic energy $E^{\rm{fluid}}$, the bosonic energy $E^{\phi}$ and the ratio between number of bosons and fermions $N^{B}/N^{F}$ of the corresponding equilibrium configuration. }
\centering 
\begin{tabular}{c c c c | c c  c c c | c c c c }
\hline
\hline                  
Model & $\rho_{c}$ & $\Lambda$ & $A_{0}$ & $\omega_{1}$ & $\omega_{2}$  & $E^{\rm{fluid}}_{30}$ & $E^{\phi}_{30}$ & $N^{B}_{30}/N^{F}_{30}$ & $\omega$ & $E^{\rm{fluid}}$ & $E^{\phi}$ & $N^{B}/N^{F}$ \\ [0.5ex]
\hline
MS1 & $1.28 \times 10^{-3}$ & 0 & $4.5\times 10^{-4}$ & 0.705 & 0.725 &  1.5330 &  0.1305 & 0.0775 & 0.695 & 1.5166 & 0.1223 & 0.0805     \\
MS2 & $1.28 \times 10^{-3}$ & 30 & $4.5\times 10^{-4}$  & 0.696 & 0.720 & 1.5380 & 0.1290 & 0.0813 & 0.715 & 1.531 & 0.1289 & 0.0839  \\
MS3 & $1.28 \times 10^{-3}$ & -30 & $4.0\times 10^{-4}$ & 0.703 & 0.729  & 1.5751 &  0.0719  & 0.0423 & 0.696 & 1.569 & 0.0696   & 0.0444    \\
MS4 & $1.28 \times 10^{-3}$ & 0  & $4.0\times 10^{-4}$ & 0.720 & 0.745  & 1.5548 & 0.0956 &  0.0496  & 0.715 & 1.556 & 0.0795 & 0.0511   \\
MS5 & $1.28 \times 10^{-3}$ & 30 & $4.0\times 10^{-4}$ & 0.731 & 0.752  & 1.5679  & 0.1053  & 0.0568 & - & - & - & -   \\
\hline
\hline
\end{tabular}
\label{table:models1}
\end{table*}
\begin{table}[t]
\caption{Initial models leading to Schwarzschild black hole formation. As no fermion-boson star forms for these models we only report the initial parameters of the bosonic cloud. Columns indicate the central rest-mass density $\rho_{c}$, the self-interaction parameter $\Lambda$, and the amplitude of the scalar field profile $A_{0}$.}
\centering 
\begin{tabular}{c c c c }
\hline
\hline                  
Model & $\rho_{c}$ & $\Lambda$ & $A_{0}$  \\ [0.5ex]
\hline
MS6 & $3.15 \times 10^{-3}$ & -30 & $3.5\times 10^{-4}$   \\
MS7 & $3.15 \times 10^{-3}$ & 0 & $3.5\times 10^{-4}$    \\
MS8 & $3.15 \times 10^{-3}$ & 30 & $3.5\times 10^{-4}$    \\
\hline
\hline
\end{tabular}
\label{table:models2}
\end{table}

As described in section \ref{sec:initial_formation} we start with an initial configuration describing a bosonic cloud of matter surrounding an already formed FS, and we study the accretion of the bosonic matter on to the FS. The bosonic cloud loses part of its energy through gravitational cooling and plunges towards the center of the FS. Intuitively, this process can lead to two possible outcomes: either to the formation of a fermion-boson star or, if the mass of the entire system is above a certain threshold, to the formation of a Schwarzschild black hole.

During the evolutions we compute useful physical quantities in order to keep track of the formation process and to evaluate the features of the final object. Those will be used below to compare with some of our static models. We define the bosonic and fermionic energy contained in spheres of different radii $r^{*}$ as
\begin{align} 
E_{r^* }^{\rm{fluid}} & = 4\pi\int_{0}^{r^*}\mathcal{E}^{\rm{fluid}}\sqrt{\gamma} dr, \label{energy_HD} \\
E_{r^* }^{\phi} & =4 \pi \int_{0}^{r^*}\mathcal{E}^{\phi}\sqrt{\gamma} dr \label{energy_scalar} ,
\end{align}
where $\sqrt{\gamma}=\psi^{6}\sqrt{a}b r^2$  is the spatial volume element for the metric \eqref{isotropic_metric}. Note that we will refer to $E^{\rm{fluid}/\phi}_{r_{\rm{max}}}$ when referring to the total energy in the computational grid. Other useful quantities we evaluate along the numerical evolution are the number of bosonic and fermionic particles within spheres of radii $r^{*}$, computed by means of the following integrals
\begin{align} 
N^{B}_{r^* } & =4\pi \int_{0}^{r^*}g^{0\nu}J_{\nu}\alpha\sqrt{\gamma} dr, \label{number_bosons} \\
N^{F}_{r^* } & =4\pi \int_{0}^{r^*} \rho \sqrt{\gamma} dr \label{number_fermions} ,
\end{align}
where $J_{\nu} = \frac{i}{2}(\bar{\phi}\,\partial_{\nu}\phi - \phi\,\partial_{\nu}\bar{\phi})$ is the conserved current associated with the transformation of the U(1) group. We also extract the scalar-field frequency $\omega$ by performing a Fast Fourier transform (FFT) of the real/imaginary part of the scalar field $\phi$. The time window for the FFT is chosen at a sufficiently  late time of the evolutions, once the bosonic cloud has already accreted on to the FS and the final object oscillates around an equilibrium configuration.

For our study we use two different FS models, both described by the polytropic EoS, $P=K \rho^{\Gamma}$, with different central value of the rest-mass density $\rho_{c}$. We consider the same scalar-field mass parameter, $\mu=1$, frequency, $\omega=0.8$, and three different values for the self-interaction parameter $\Lambda = \{-30,0,+30\}$. Our model for the bosonic cloud, equation~\eqref{scalarfield}, has a couple of free parameters we can vary, namely the amplitude $A_{0}$ and the width of the Gaussian profile $\sigma$. For all our models we consider $\sigma=90$ which corresponds to a bosonic cloud much larger than the FS radii. We summarize some of the properties of our initial models in Table~\ref{table:models1} and Table~\ref{table:models2}.

\begin{figure}[t!]
\begin{minipage}{1\linewidth}
\includegraphics[width=0.95\textwidth]{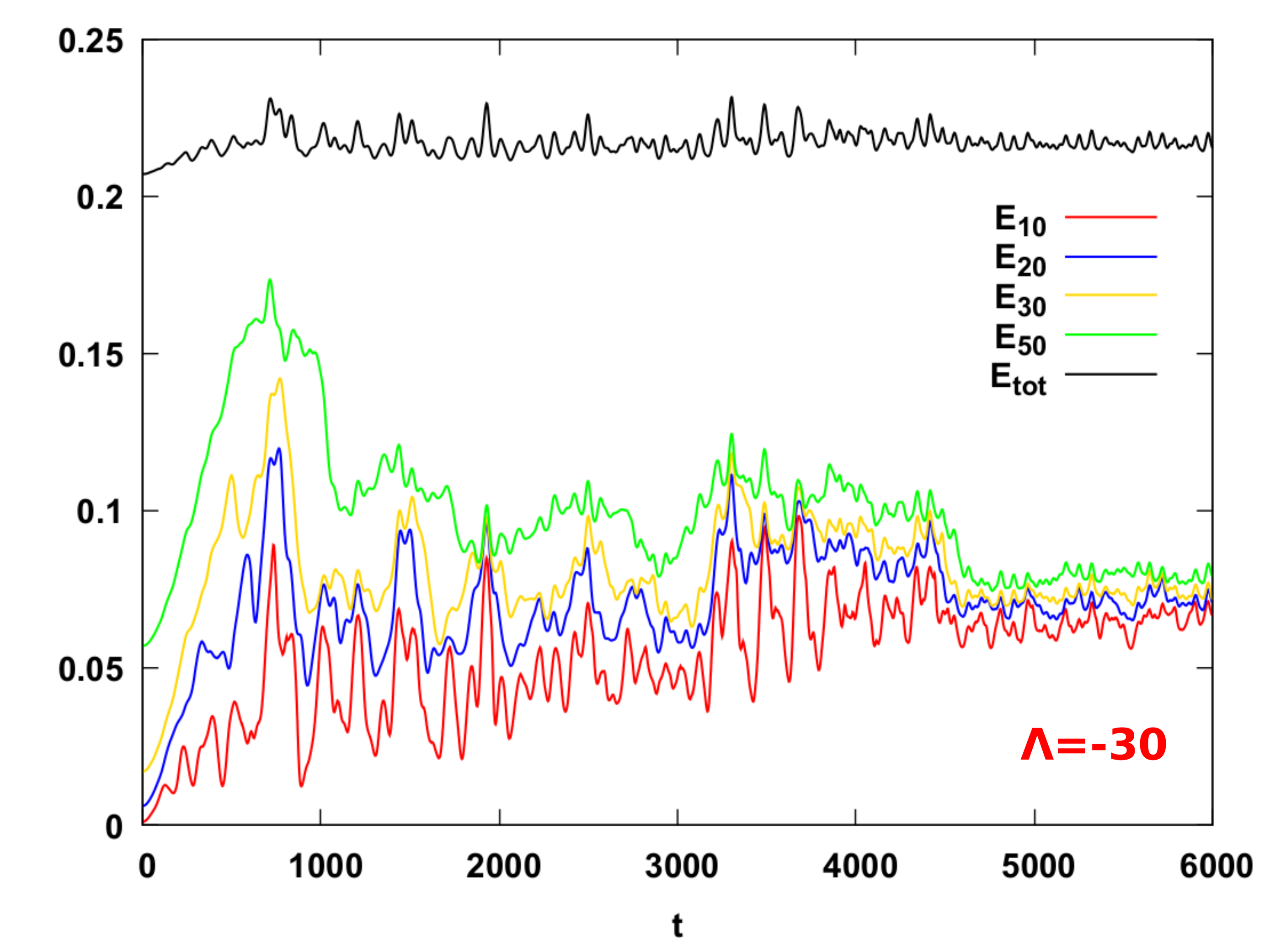}  
\includegraphics[width=0.95\textwidth]{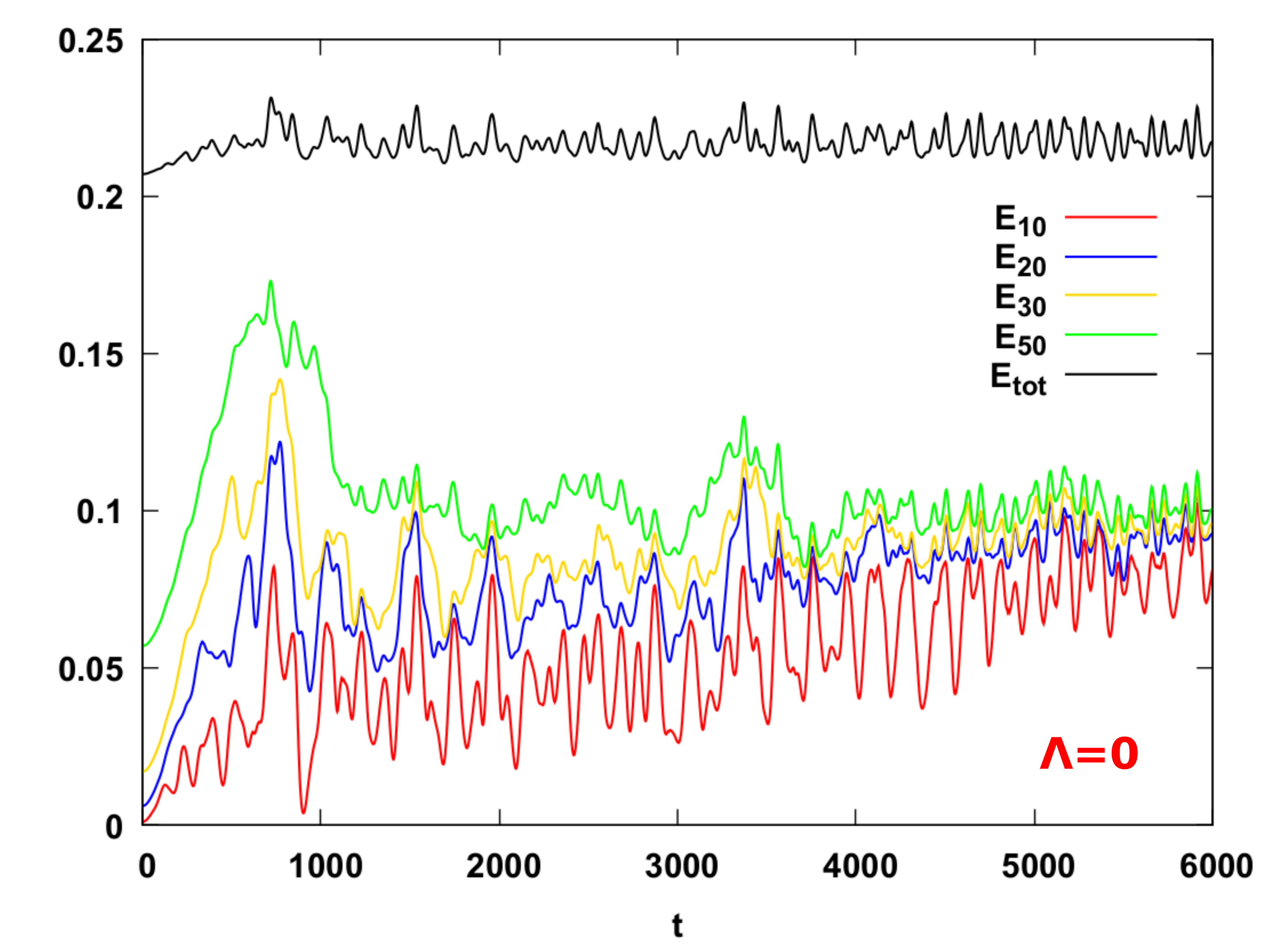}  
\includegraphics[width=0.95\textwidth]{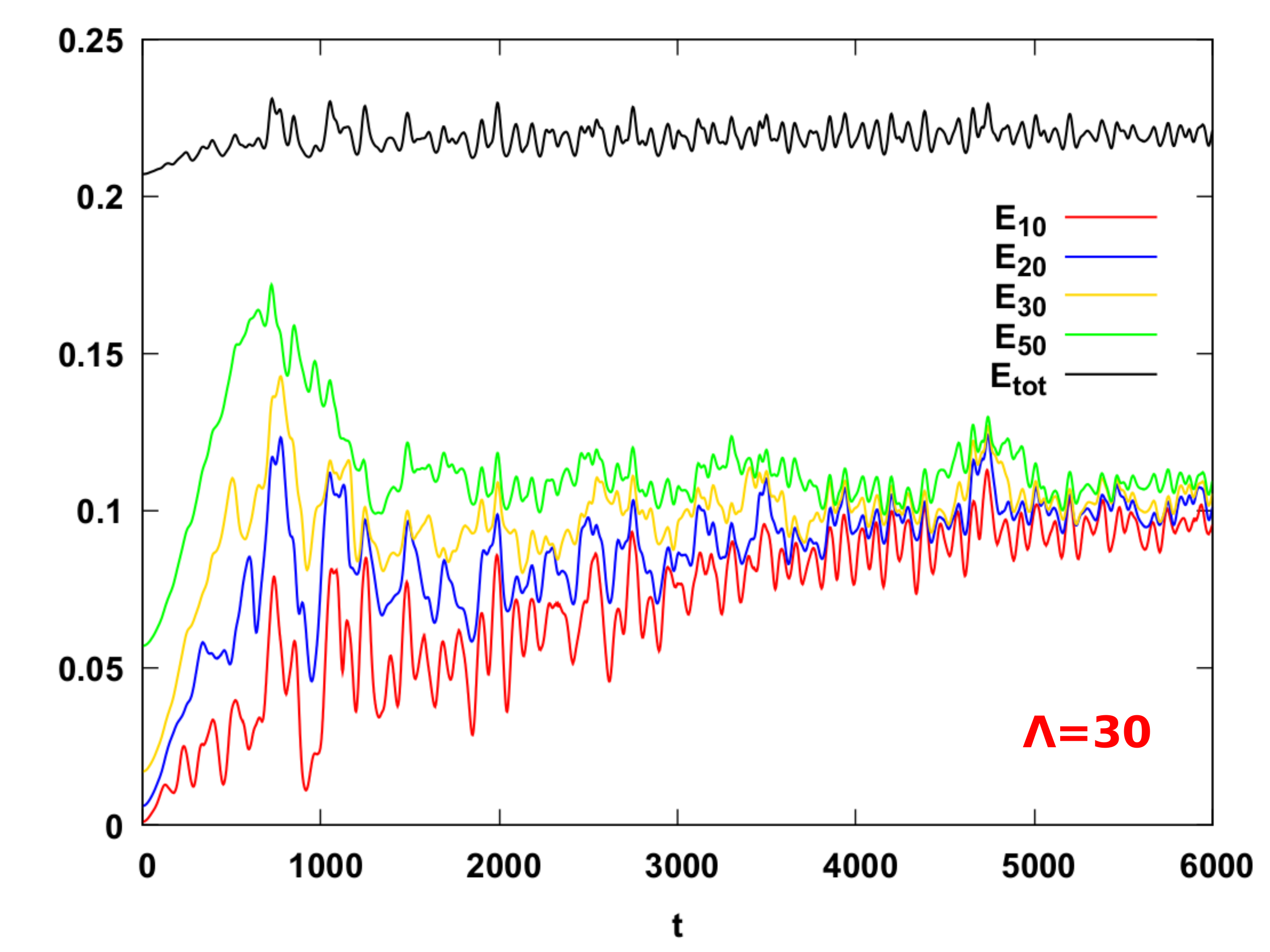}  
\caption{Scalar-field energy in spheres of different radii, for model MS3 (top), MS4 (center), and MS5 (bottom). The red, blue, gold and green lines, correspond to $r^*=10, 20, 30, 50$, respectively. The black line represents the total energy.}
\label{fig:energies_MS5}
\end{minipage}
\end{figure}

In Fig.~\ref{fig:energies_MS5} we show the evolution of the scalar-field energy contained in spheres of different radii $r^{*}$ calculated with Eq.~\eqref{energy_scalar}, for models MS3, MS4 and MS5 described in Table~\ref{table:models1}. The growth of the lines $E^{\phi}_{50}$, $E^{\phi}_{30}$, $E^{\phi}_{20}$, and $E^{\phi}_{10}$ shows that during the evolution the energy of the scalar cloud, which at the initial time is spread over a large spatial volume, gradually concentrates in a smaller volume, as it is being accreted by the FS. Part of the cloud energy does not fall on to the FS but it is radiated away through the gravitational cooling mechanism. For all three models, from time $t\simeq 750$ the curves start to converge slowly to each other, indicating that the scalar field is contained within small radii, radiating the excess energy to infinity. The remnant energy is confined into a volume delimited by $r\simeq 6$ and is, hence, entirely contained inside the FS. The total scalar-field energy of the three models is $E^{\phi}\simeq 0.065$ for MS3, $E^{\phi}\simeq 0.09$ for MS4, and  $E^{\phi}\simeq 0.11$ for MS5. 

Figure~\ref{fig:energies_MS5} shows some differences between models with and without self-interaction, and also depending on the sign of the self-interaction term. In the case with $\Lambda=-30$ (top panel) the lines $E^{\phi}_{50}$, $E^{\phi}_{30}$, $E^{\phi}_{20}$ and $E^{\phi}_{10}$ slowly converge to each other and, at around $t\simeq 4700$, the energy within larger volumes radiates away and all the lines converge to the red one ($E^{\phi}_{10}$) with a final energy around $E^{\phi}\simeq 0.065$. 
For the case with positive $\Lambda$ (bottom panel), again there is an initial phase during which the lines slowly converge to each other, but then the red line, corresponding to $E^{\phi}_{10}$, grows reaching the green one, $E^{\phi}_{50}$. This indicates that all the scalar matter around the forming compact object is accreting onto it. The case with $\Lambda=0$ (central panel) is an intermediate case, with the lines slowly converging to each other for the entire evolution. This result can be understood as follows: a $\Lambda>0$ term in the Lagrangian is an attractive term, helping gravity letting the cloud collapse on to the FS and acting against the gravitational cooling mechanism that radiates away scalar particles. This means that the formation process is accelerated and the final object will also have higher number of bosonic particles and mass. On the other hand, $\Lambda<0$ is a repulsive term, which increases the amount of bosonic particles expelled to infinity. Nonetheless the formation process seems to be accelerated but it is due to the fact that the scalar particles around the formed compact object escape faster to infinity. We point out that, as $|\phi|<1$, the self-interaction term, which is proportional to $\lambda|\phi|^{4}$, gives a lower order contribution than the mass term $\mu^{2}|\phi|^{2}$. They are only comparable when the object is compact enough to reach high values of $\phi$. This is the reason why the first part of the evolution before the object forms is basically the same for the three models.

\begin{figure}[t!]
\begin{minipage}{1\linewidth}
\includegraphics[width=0.95\textwidth]{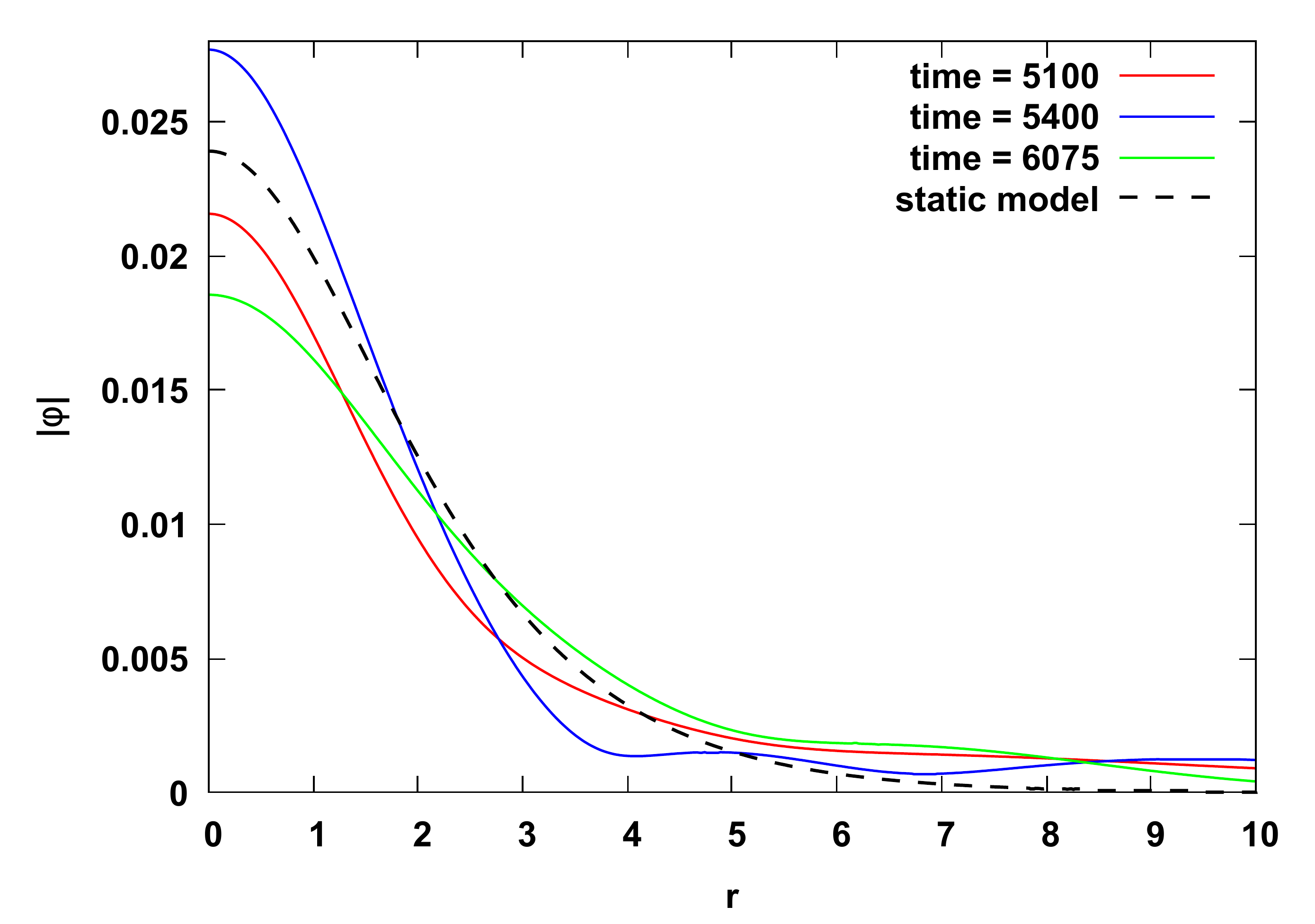}  \\
\includegraphics[width=0.95\textwidth]{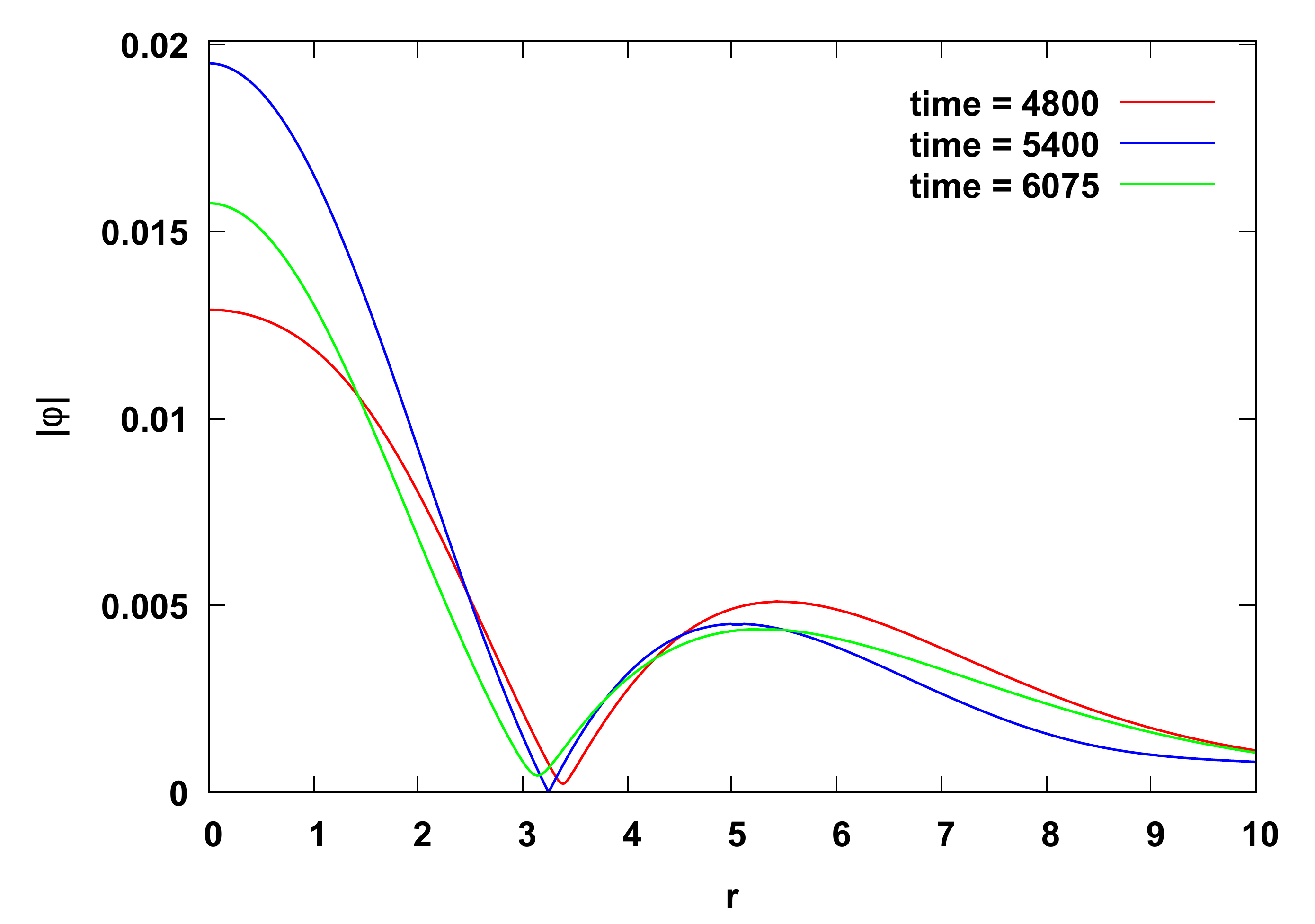}  
\caption{Late-time radial profiles of the scalar field module $|\phi|$ for model MS3 with $\Lambda=-30$ (upper panel) and model MS5 with $\Lambda=30$ (bottom panel). The three snapshots of model MS3 are compared with the radial profile of a static mixed star model of similar $\rho_{c}$, $\phi_{c}$, and bosonic and fermionic energy and number (dashed black line in the plot). Model MS5 presents a node at $r\simeq 3$ that radially oscillates together with the rest of the profile.}
\label{fig:MS6_latetimephi}
\end{minipage}
\end{figure}

In Fig.~\ref{fig:MS6_latetimephi} we depict the late-time radial profiles of the scalar field module $|\phi|$  for models MS3 (top panel) and MS5 (bottom panel). For model MS3 we compare three different snapshots during the evolution with an equilibrium configuration of a mixed star with comparable mass and number of bosons and fermions. The comparison shows  that the radial profile of $|\phi|$ obtained through the dynamical formation process resembles that of the static solution. 

The bottom panel of Fig.~\ref{fig:MS6_latetimephi} shows that for model MS5 there are two maxima of the scalar field and there is a node at around $r\simeq 3$ which oscillates radially with the rest of the profile. At first sight this result seems surprising because, at least for boson stars, all models with nodes are in excited states, which are intrinsically unstable and collapse to a black hole or decay to the nodeless fundamental configuration~\cite{Balakrishna:1997ej,Lee:1988av}. We note that in~\cite{Bernal2010} configurations of two coexisting states of the scalar field, the ground state and one excited state, were investigated. Their results showed that it is possible to combine an intrinsically unstable first excited state (with a node) and the ground nodeless configuration, and obtain a stable configuration. In the fermion-boson case analysed here, our results seem to indicate that an excited state of the scalar field in the presence of fermionic matter may form a stable configuration as well.

\begin{figure}[t!]
\begin{minipage}{1\linewidth}
\includegraphics[width=0.95\textwidth]{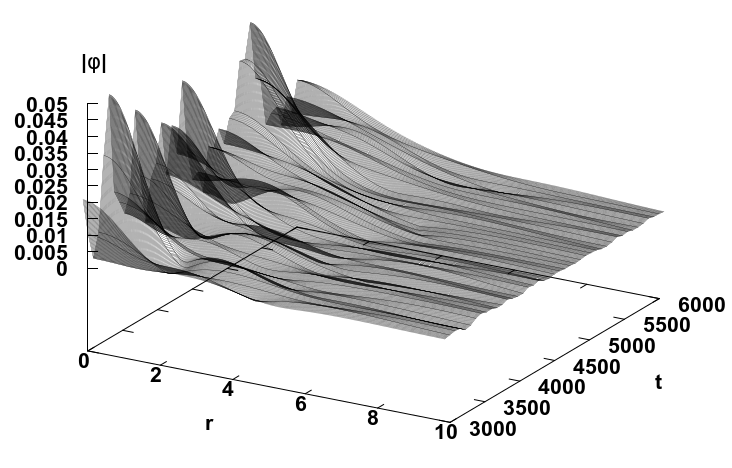}  \\
\includegraphics[width=0.95\textwidth]{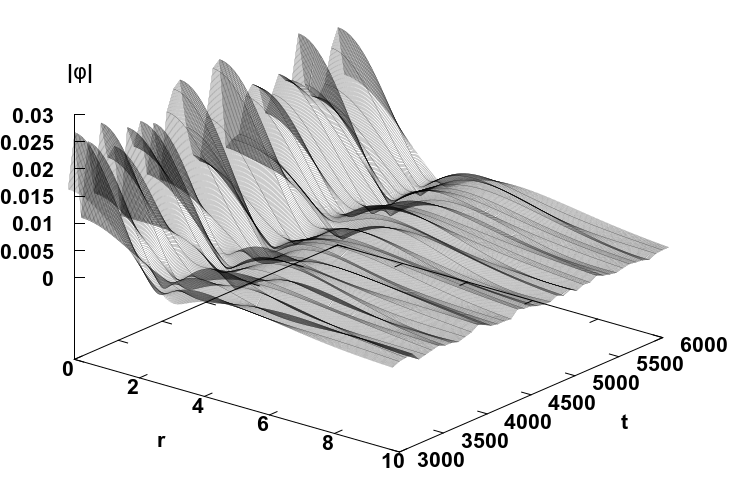}  
\caption{ Evolution of the radial profile of the scalar field module $|\phi|$ for model MS3 with $\Lambda=-30$ (upper panel) and model MS5 with $\Lambda=30$ (bottom panel)  in the time window $t \in [3000,6000]$. 
The difference between configuration MS3, which does not show a node in the last part of the evolution, and the excited state MS5 is apparent.}
\label{fig:3dplots}
\end{minipage}
\end{figure}

To provide further grounds for this result, Fig.~\ref{fig:3dplots} depicts a 3D plot of the late-time evolution of the scalar field for models MS3 and MS5. The presence of the node for model MS5 (bottom plot) is clearly visible. This figure shows that this is not just a transient state as the evolution is characterised by radial oscillations around an equilibrium configuration. This is in contrast with model MS3 where we can only see transient nodes in the scalar profile which are due to the bosonic particles radiated away through gravitational cooling. This and previous results~\cite{Bernal2010} would indicate that mixed states that only interact through gravity and in which one of the components is intrinsically unstable, can cooperate so they become globally stable.

\subsection{Evolutions of the equilibrium configurations} \label{sec:evolutions}

In Section~\ref{stat} we discussed how we identify the region of the parameter space where stable configurations are found. In this section we intend to verify the results obtained by performing numerical evolutions of stable and unstable models. We expect stable mixed stars to show a combination of the typical behaviour of isolated stable boson stars and fermion stars. This means that we expect the scalar field to oscillate with its characteristic eigenfrequency $\omega$ while the fermionic density $\rho$ is expected to oscillate slightly around its initial state due to the numerical truncation errors introduced by the discretization of the differential equations of the continuum model. All physical quantities of the stable models such as  mass, boson number density or fermion number density are expected to be constant in time. Even under the introduction of a small perturbation, stable models are expected to oscillate around their static solutions.

For a model in the unstable region, however, we expect the small-amplitude perturbations induced by the numerical errors to grow due to the non-linearity of the system. The growth of the perturbations can lead to three different outcomes: the migration to the stable region, the gravitational collapse and formation a Schwarzschild black hole, or the dispersion of the bosonic particles. 

\begin{figure*}[t!]
\centering
\includegraphics[width=0.33\textwidth]{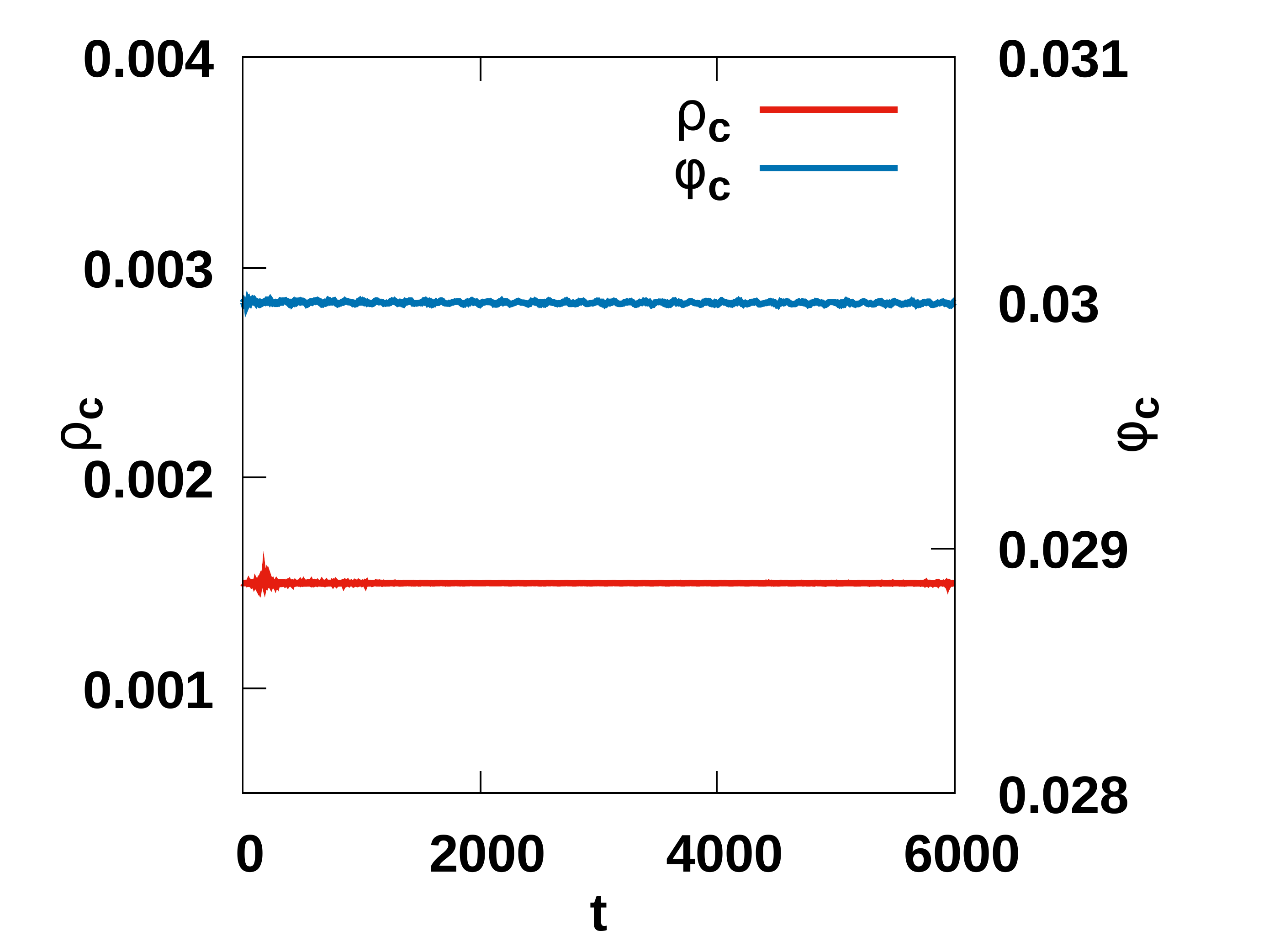} \hspace{-0.2cm}
\includegraphics[width=0.33\textwidth]{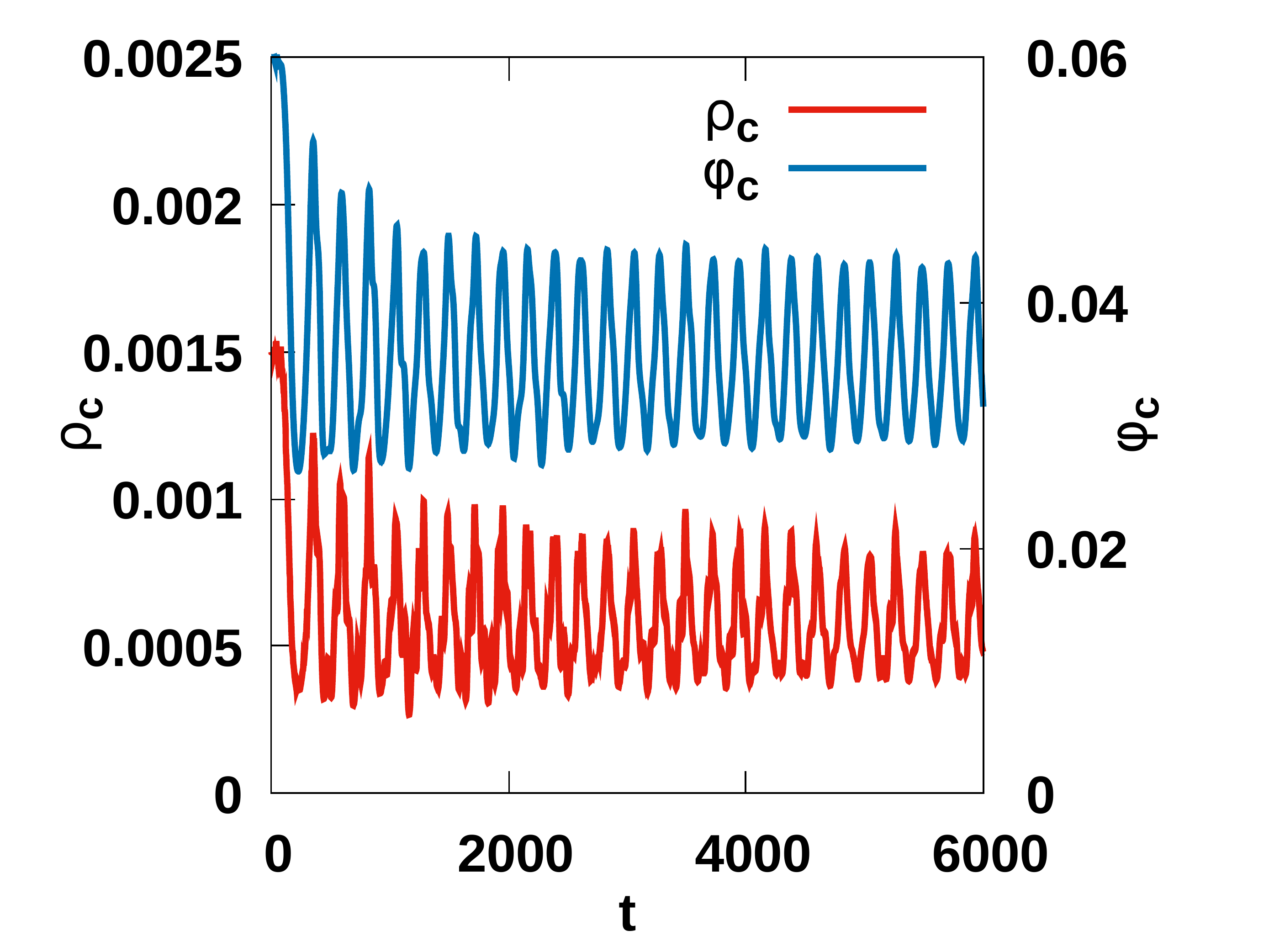} \hspace{-0.07cm}
\includegraphics[width=0.33\textwidth]{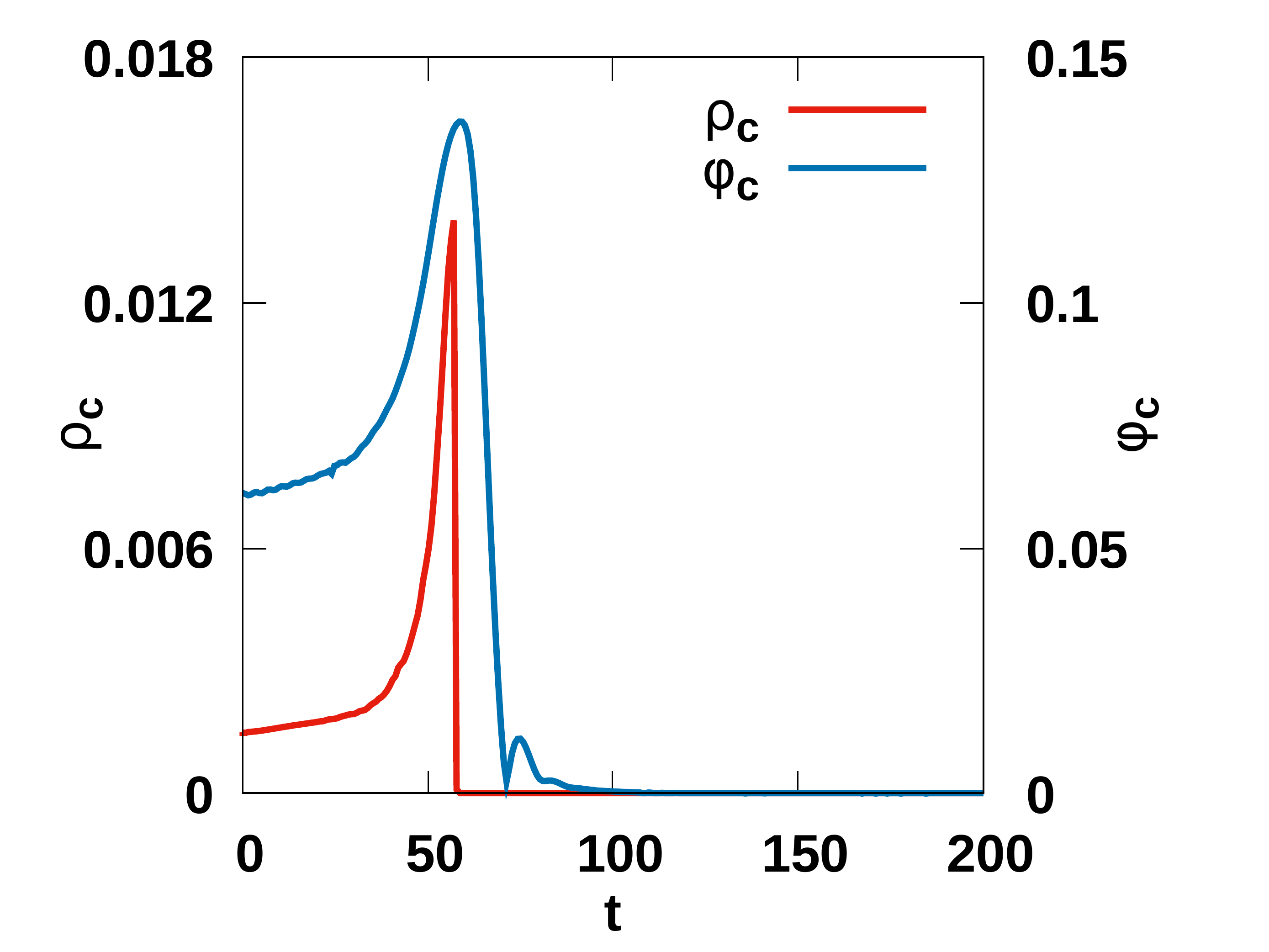} \\
\includegraphics[width=0.33\textwidth]{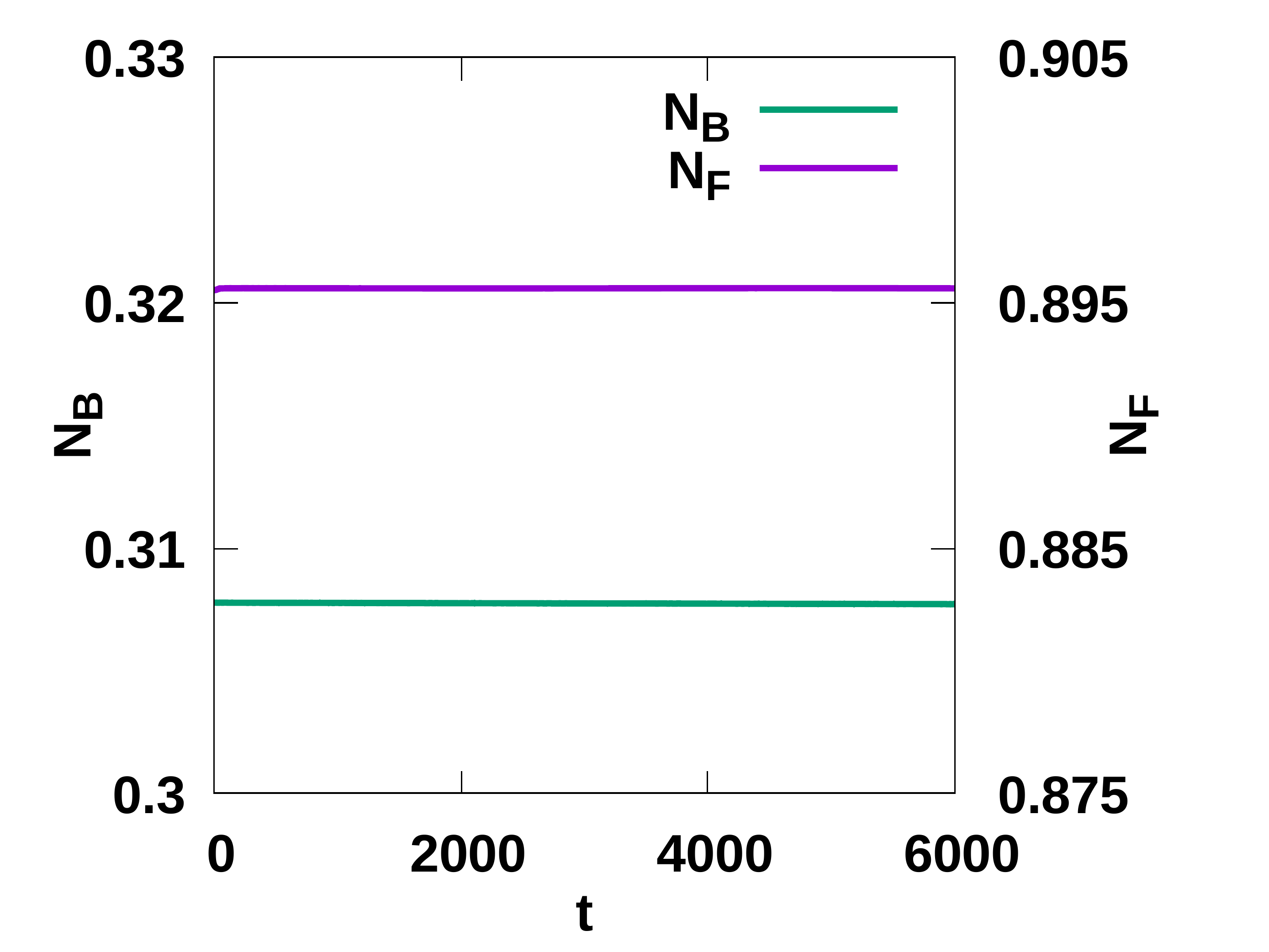} \hspace{0.1cm}   
\includegraphics[width=0.33\textwidth]{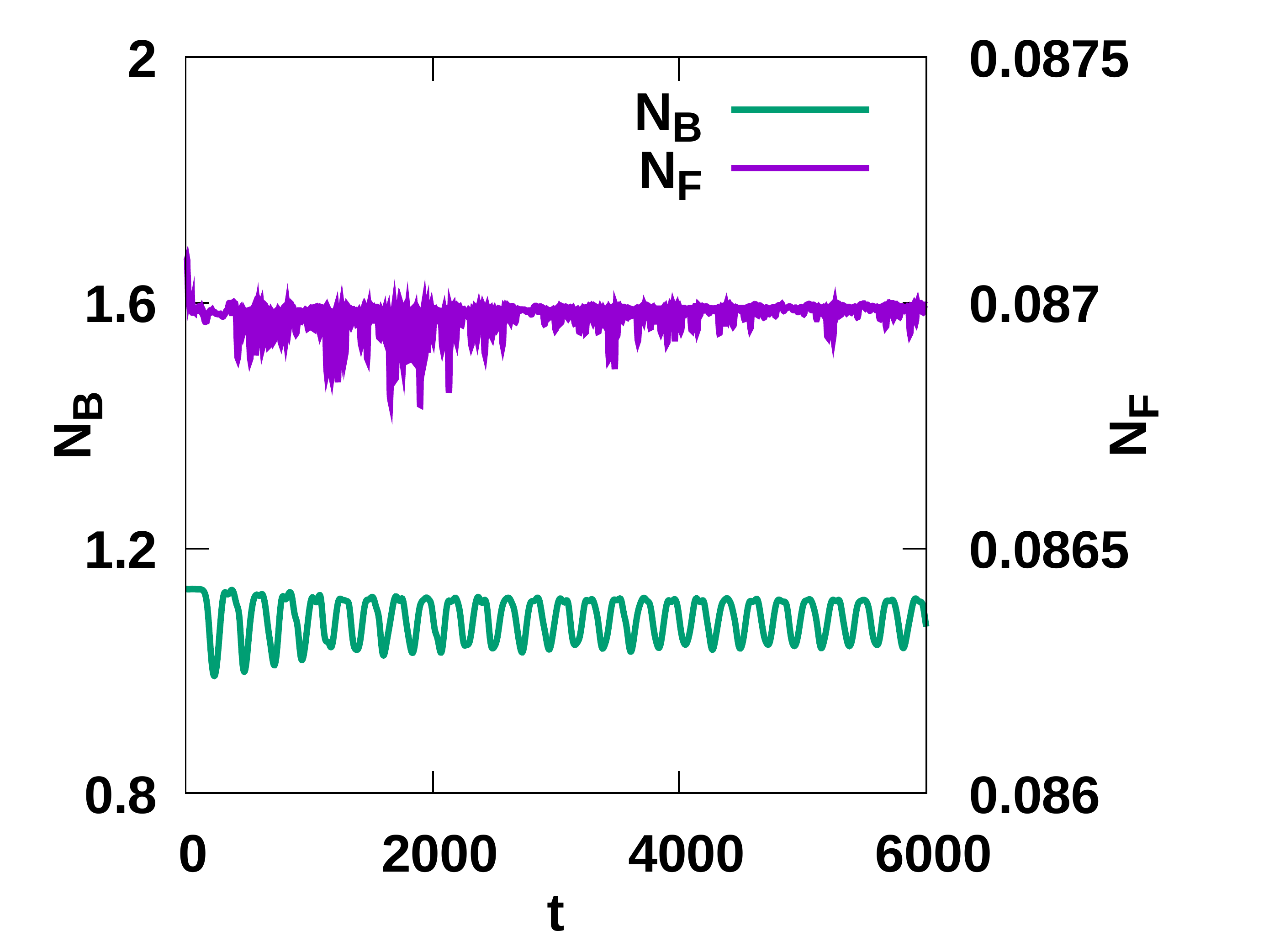} \hspace{-0.1cm}
\includegraphics[width=0.27\textwidth]{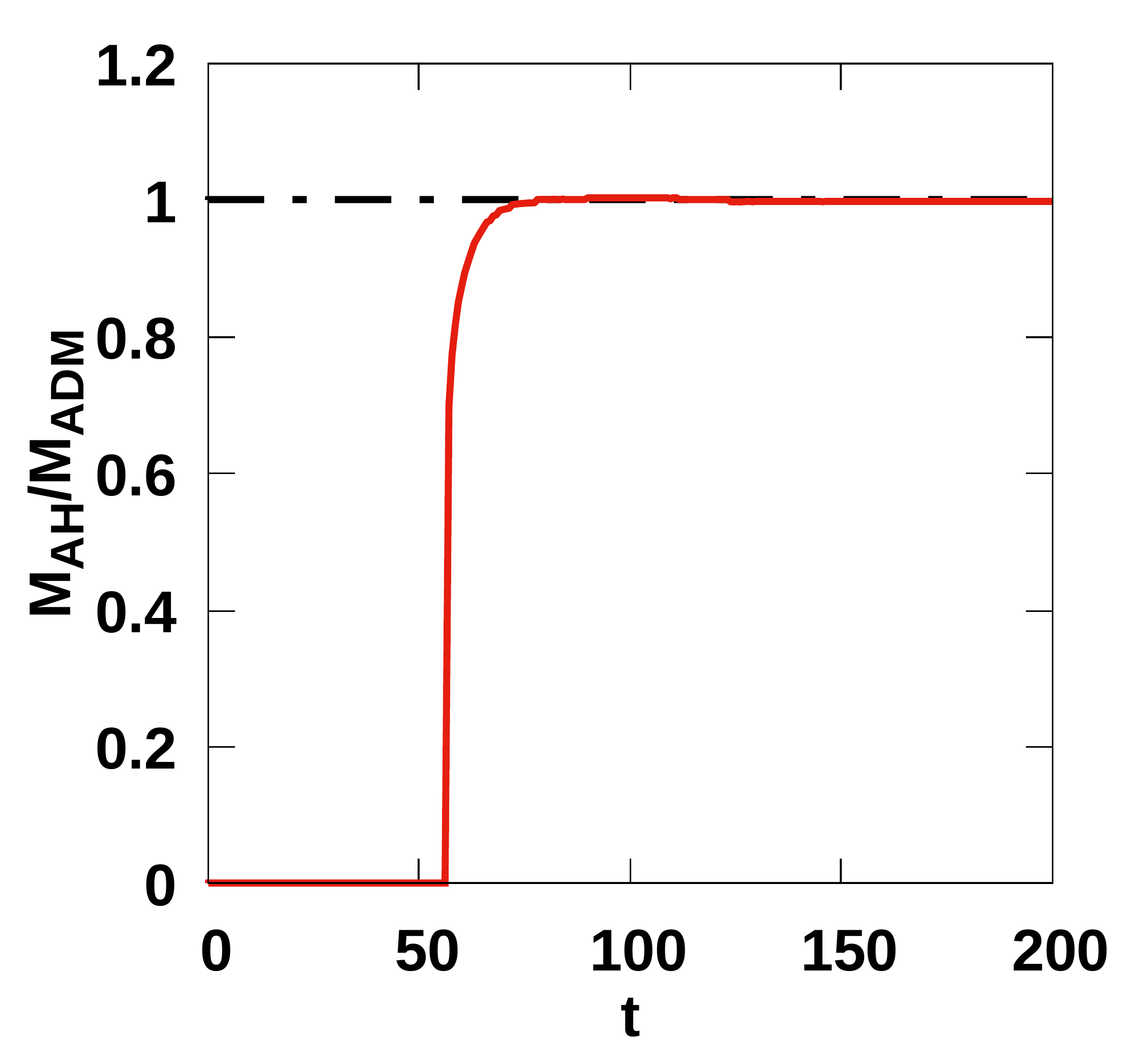}\hspace{0.7cm}

\caption{Time evolution of static models with self-interaction parameter $\Lambda = 30$. Left panels depict the central value of the fluid density $\rho_{c}$ and of the scalar field $\phi_{c}$ (top row) and number of bosons $N_{B}$ and fermions $N_{F}$ (bottom row) for the stable model MS11. Middle panels show the same physical quantities for the unstable model MS12 without the addition of an artificial perturbation. The right panels show the collapse to a Schwarzschild black hole of model MS12 when a  $2\%$ perturbation is induced in the scalar field. The right bottom plot displays the AH mass in units of the ADM mass (red solid line) and the time evolution of the ADM mass normalized by its initial value (black dashed line).}
\label{fig6}
\end{figure*} 

\begin{table*}[t]
\caption{Static fermion-boson star models. From left to right the columns indicate the model name, its stability, the value of the self-interaction parameter $\Lambda$, the central value of the fluid density $\rho_c$ and of the scalar field $\phi_c$, the field frequency obtained with the shooting method $\omega_{\rm{shoot}}$, the normalized frequency $\omega$, the number of bosons to fermions ratio $N_{B}/N_{F}$, the number of bosons $N_{B}$, the radius containing $99\%$ of bosons, fermions and total particles, $R_{B}$, $R_{F}$, $R_{T}$, respectively. All radii are evaluated using Schwarzschild coordinates.}
\centering 
\begin{tabular}{c  c | c c  c c c | c c c c c}
\hline
\hline                  
Model & Branch & $\Lambda$ & $\rho_{c}$  & $\phi_c$ & $\omega_{\rm{shoot}}$ & $\omega$ & $N_{B}/N_{F}$ & $N_{B}$ & $R_{B}$ & $R_{F}$ & $R_{T}$ \\ [0.5ex]
\hline
MS9 & Stable & $0$ & $1.88\times 10^{-3}$ & $4.04\times 10^{-2}$ & $1.199$ & $0.732$ & $0.191$ & $0.208$ & $5.51$ & $7.22$ & $7.08$ \\
MS10 & Unstable & $0$ & $4.55\times 10^{-3}$ & $8.03\times 10^{-2}$ & $1.436$ &  $0.602$ & $0.251$ & $0.261$ & $3.66$ & $5.81$ & $5.63$ \\

MS11 & Stable & $30$ & $1.50\times 10^{-3}$ & $3.00\times 10^{-2}$ & $1.238$ & $0.807$ & $0.344$ & $0.308$ & $7.06$ & $7.08$ & $7.04$ \\
MS12 & Unstable & $30$ & $1.50\times 10^{-3}$ & $6.00\times 10^{-2}$ & $1.629$ &  $0.815$ & $13.03$ & $1.134$ & $6.93$ & $3.06$ & $6.81$ \\

MS13 & Stable & -30 & $1.83\times 10^{-3}$ & $3.02\times 10^{-2}$ & $1.129$ & $0.680$ & $0.055$ & $0.079$ & $5.18$ & $7.73$ & $7.64$ \\
MS14 & Unstable & -30 & $2.41\times 10^{-3}$ & $6.06\times 10^{-2}$ & $1.099$ &  $0.603$ & $0.068$ & $0.097$ & $3.99$ & $7.34$ & $7.24$ \\

\hline
\hline
\end{tabular}
\label{table3}
\end{table*}

As our evolution code is based on isotropic coordinates \eqref{isotropic_metric} and the mixed-star models are constructed using Schwarzschild coordinates \eqref{Schwarzschild_metric}, we must apply a coordinate transformation to be able to evolve the initial configurations. We follow the procedure proposed in~\cite{Kleihaus1998} which can be divided in two steps. First, we perform the change of coordinates noting that from the comparison between the two metrics we have that
\begin{equation}\label{coord_change1}
\frac{d\hat{r}}{dr} = \tilde{a}(r) \frac{\hat{r}}{r}\,.
\end{equation}
To obtain the coordinate transformation we introduce the function $\beta$ 
\begin{equation}
\beta = \frac{\hat{r}}{r}.
\end{equation}
Rewriting equation \eqref{coord_change1} in terms of $\ln\beta$ we obtain
\begin{equation}
\frac{d\ln\beta}{dr} = \frac{1}{r} ( \tilde{a}(r) -1 ),
\end{equation}
which leads to
\begin{equation} \label{integral_beta}
\beta(r) = \exp{\left[ -\int_{r}^{r_{\rm max}} \frac{1}{r'} \left(\tilde{a}(r')-1\right) dr'\right]}\,.
\end{equation}
As initial condition to solve this integral, we impose that at the outer boundary the spacetime resembles the Schwarzschild solution which yields
\begin{equation}
\hat{r}_{\rm max} = \left( \frac{1+\sqrt{\tilde{a}(r_{\rm max})}}{2}\right)^2 \frac{r_{\rm max}}{\tilde{a}(r_{\rm max})}\,.
\end{equation}
Once we obtain $\beta$, we can finally obtain the conformal factor which is defined as
\begin{equation}
\psi = \sqrt{\frac{r}{\hat{r}}} = \sqrt{\frac{1}{\beta}}.
\end{equation}
We point out that the introduction of the new variable $\beta$ is necessary to make the integral~\eqref{integral_beta} behave well at the origin, and to be able to reconstruct the solution in the entire radial domain. The interested reader is addressed to~\cite{Kleihaus1998} for further details.

We perform evolutions of several models for values of the self-interaction parameter $\Lambda = \{-30,0,30\}$, both in the stable and unstable region of the existence surface. These numerical evolutions confirm our analysis about the stability of the models. We summarize their relevant physical properties in Table~\ref{table3}.

Figure~\ref{fig6} shows the time evolution of the results obtained for the case $\Lambda=30$, in particular models MS11 and MS12 of Table~\ref{table3}. In the left panels we display the evolution of the central value of the fluid density $\rho_c$ and of the scalar field $\phi_c$ (top row) and the evolution of the number of fermions and bosons (bottom row), for the stable model MS11. As expected all these physical quantities remain constant in time confirming that the model is stable. The middle panels show the time evolution of the same physical quantities for model MS12, which is in the unstable region. We can observe that the central values of the scalar field and the fluid density very rapidly depart from their initial values, with a large variation which is damped in a few cycles. The system settles on a new configuration in the stable branch, oscillating around the new central values $\rho_{c} \simeq 0.0007$ and $\phi_{c} \simeq 0.038$. The number of bosons and fermions oscillate around a value very close to the initial one. These results indicate that this unstable model is migrating to a new configuration in the stable branch. 

Finally, in the right panels of Fig.~\ref{fig6} we show the evolution of the same model MS12 under the effects of a perturbation. To do so we replace the initial profile of the scalar field with 
\begin{align}
\phi(r) \rightarrow \phi(r) \left(1 + \frac{A_{1}}{100}\right),
\end{align}
where $A_{1}=2$, which corresponds to a $2\%$ level perturbation. Despite fairly small, this artificial perturbation is stronger than that introduced by the discretization errors alone which triggered the evolution shown in the middle panels of Fig.~\ref{fig6}. We now observe that due to the stronger perturbation the model does not migrate to the stable region but rather collapses to a Schwarzschild black hole, as signalled by the formation of an apparent horizon (AH). In the top row we show the time evolution of the central values of the fluid density and of the scalar field while in the bottom row we show the time evolution of the mass of the black hole evaluated on the AH in units of the ADM mass of the system (which we depict with a dashed black curve). We could not find any model for which the bosonic part dispersed, leaving behind a purely FS. The binding energy of the whole configuration is never positive and therefore, unstable models can only either migrate or collapse.

\section{Conclusions}
\label{sec:conclusions}

Fermion-boson stars are gravitationally bound structures composed by fermions and scalar particles.
They are regular and static macroscopic configurations obtained 
by solving the coupled Einstein-Klein-Gordon-Euler system. In this paper we 
have discussed a possible scenario through which fermion-boson stars may 
form assuming an initial configuration in which an already existing FS (i.e.~a neutron star)
is surrounded by an accreting dilute cloud (a Gaussian pulse) of a massive, complex scalar field. Our setup has considered positive and negative values of a quartic self-interaction term in the Klein-Gordon potential. 
We have built constraint-satisfying initial data and we have modelled the astrophysical situation by considering different bosonic cloud amplitudes and widths and two different fermion star models.
The results of our spherically-symmetric, numerical-relativity 
simulations have shown that once part of the initial scalar field is expelled via gravitational 
cooling the system oscillates around an equilibrium configuration that is asymptotically 
consistent with the static solutions of the system. 

Existence diagrams of such equilibrium solutions in the central-field-amplitude vs central-fermionic-density plane have been constructed to draw such comparisons. Our results are in agreement, in the corresponding limits, with the work of~\cite{valdez2013dynamical,valdez2020fermion}.
The non-linear stability of static models residing in both the stable and unstable regions of the existence diagrams has been assessed  through simulations with a quartic self-interaction potential in the bosonic sector, not attempted  in  previous works. Those have shown that, for stable configurations, all physical quantities describing the star, such as energy and number of particles, remain constant during the evolution, while unstable models either migrate to the stable region or  collapse to a Schwarzschild black hole.

The dynamical formation of fermion-boson stars for large positive values of the coupling constant in the quartic self-interaction term (namely $\Lambda=30$) has revealed the presence of a node in the scalar field. This is an intriguing result as purely boson stars with nodes correspond to excited states and are known to be intrinsically unstable~\cite{Balakrishna:1997ej,Lee:1988av}. However, fermion-boson stars with nodes in the bosonic sector can dynamically form and appear long-term stable. This indicates that an excited state of the scalar field in the presence of fermionic matter may form a stable configuration. This result is akin to the findings of~\cite{Bernal2010} who found that boson star configurations in which the ground state and the first excited state of the scalar field coexist are stable. In upcoming investigations we plan to build equilibrium fermion-boson configurations with an excited state of the scalar field and study their stability properties to confirm the result reported here. Likewise, we will analyze the dynamical formation of rotating mixed stars as it might as well be possible that the presence of fermionic matter stabilized otherwise unstable spinning boson stars~\cite{sanchis2019nonlinear}.


\acknowledgments
We thank Eugen Radu and Carlos Herdeiro for useful suggestions. This work was supported by the Spanish Agencia Estatal de Investigaci\'on (grant PGC2018-095984-B-I00), by the Generalitat Valenciana (PROMETEO/2019/071 and GRISOLIAP/2019/029), by the European Union’s Horizon 2020 RISE programme H2020-MSCA-RISE-2017 Grant No.~FunFiCO-777740, by DGAPA-UNAM through grants No.~IN110218, IA103616, IN105920, by the Funda\c c\~ao para a Ci\^encia e a Tecnologia (FCT) projects PTDC/FIS-OUT/28407/2017 and UID/FIS/00099/2020 (CENTRA), and CERN/FIS-PAR/0027/2019. SF gratefully acknowledges support by the Erasmus+ International Credit Mobility Program KA-107 for an academic stay at the University of Valencia.


\bigskip


\bibliography{num-rel2}

\end{document}